%% file: main.tex
\documentclass[twocolumn]{aastex631}
\usepackage{natbib}
\usepackage{amsmath}

\newcommand{\teff}[1]{$T_{\rm eff}$#1}
\newcommand{\Prot}[1]{$P_{\rm rot}$#1}
\newcommand{\logrhk}[1]{$\log{R'_{\text{HK}}}$#1}
\newcommand{\vsini}[1]{$v \text{sin} i$#1}
\newcommand{\name}[1]{TOI-1695#1}

\begin{document}

\suppressAffiliations
\input{authors}

\correspondingauthor{Collin Cherubim}
\email{ccherubim@g.harvard.edu}

\title{\textbf{TOI-1695 b: A Water World Orbiting an Early M Dwarf in the Planet Radius Valley}}

\begin{abstract}
Characterizing the bulk compositions of transiting exoplanets within the M dwarf radius valley offers a unique means to establish whether the radius valley emerges from an atmospheric mass loss process or is imprinted by planet formation itself. We present the confirmation of such a planet orbiting an early M dwarf ($T_{\rm mag} = 11.0294 \pm 0.0074, M_s = 0.513 \pm 0.012\ M_\odot, R_s = 0.515 \pm 0.015\ R_\odot, T_{\rm eff} =3690\pm 50 K$): TOI-1695 b ($P = 3.13$ days, $R_p = 1.90^{+0.16}_{-0.14}\ R_\oplus$). TOI-1695 b's radius and orbital period situate the planet between model predictions from thermally-driven mass loss versus gas depleted formation, offering an important test case for radius valley emergence models around early M dwarfs. We confirm the planetary nature of TOI-1695 b based on five sectors of TESS data and a suite of follow-up observations including 49 precise radial velocity measurements taken with the HARPS-N spectrograph. We measure a planetary mass of $6.36 \pm 1.00\ M_\oplus$, which reveals that TOI-1695 b is inconsistent with a purely terrestrial composition of iron and magnesium silicate, and instead is likely a water-rich planet. Our finding that TOI-1695 b is not terrestrial is inconsistent with the planetary system being sculpted by thermally driven mass loss. We present a statistical analysis of seven well-characterized planets within the M dwarf radius valley demonstrating that a thermally-driven mass loss scenario is unlikely to explain this population. 

\end{abstract}

\section{Introduction} \label{introduction}

One of the most perplexing mysteries in current exoplanet science is the observed dearth of planets between 1.7 and 1.9 $R_\oplus$ around Sun-like stars \citep[$T_{\text{eff}} > 4700 K$;][]{Fulton_2017, Fulton_2018, Mayo_2018, Eylen_2018, Berger_2020} and between 1.5 and 1.7 $R_\oplus$ around mid-K to mid-M dwarfs \citep[$T_{\text{eff}} < 4700 K$;][]{Cloutier_2020a}. This so-called radius valley is most commonly thought to delineate two broad populations of planets: terrestrials and enveloped terrestrials, the latter likely possessing an extended H/He envelope and/or a volatile component such as water. The location of the rocky/enveloped transition in radius space is known to be period dependent \citep[e.g.][]{Fulton_2017,Eylen_2018, Martinez_2019, Berger_2020,Cloutier_2020a}. Distinct slopes of the radius valley in radius-period space are predicted by physical models that describe prospective pathways for the emergence of the radius valley and therefore offer a means to distinguish between different emergence models. 

Photoevaporation, core-powered mass loss, and terrestrial planet formation in a gas poor (but not gas depleted) environment predict that $R_{p, \text{valley}} \propto P^{\beta}$ where $\beta \in [-0.15, -0.09]$ \citep{Lopez_2018, Gupta_2020, Lee_2021, Rogers_2021}. Conversely, a gas depleted formation scenario in which the two populations of planets form on distinct timescales predicts a slope with the opposite sign (i.e. $\beta = 0.11$) \citep{Lopez_2018}. These model-predicted slopes carve out a wedge in period-radius space in which thermally-driven mass loss (i.e. photoevaporation and core-powered mass loss) and gas poor formation models predict that the so-called ``keystone planets'' situated within this wedge should be pure terrestrials. Conversely, the gas depleted formation model argues that they are more likely to be enveloped terrestrials because they are larger than the maximum rocky planet that can form out of the minimum mass extrasolar nebula. We define ``enveloped terrestrial'' as a rocky planet with a low-density component (e.g. H/He, volatiles) sufficient to affect its observed radius.

While numerous planet occurrence rate studies around FGK stars have shown that $-0.11\leq \beta \leq -0.09$ \citep{Fulton_2017,Eylen_2018,Martinez_2019,Rogers_2021}, which is consistent with thermally-driven mass loss or a gas poor formation mechanism, there is suggestive evidence that the slope may differ substantially around the lower mass late-K to mid-M dwarfs \citep[$\beta =0.06\pm 0.02$;][]{Cloutier_2020a}. The interpretation of this inconsistency is that around increasingly low mass stars, we may be witnessing the emergence of a new channel of terrestrial planet formation that is not strongly influenced by atmospheric escape because the terrestrial planet population that we observe today around low mass stars never accreted primordial H/He envelopes. Thus, characterizing the bulk compositions of keystone planets around M dwarfs can provide a unique observational test of this hypothesis. Such experiments are highly complementary to occurrence rate studies, which are comparatively much more time intensive.

Since 2018, NASA's Transiting Exoplanet Survey Satellite (TESS) has discovered several keystone planets around M dwarfs (TOI-776 b: \citealt{Luque_2021}; TOI-1235 b: \citealt{Bluhm_2020}, \citealt{Cloutier_2020b}; TOI-1452 b: \citealt{Cadieux_2022}, TOI-1634 b: \citealt{Cloutier_2021a}, \citealt{Hirano_2021}; TOI-1685 b: \citealt{Bluhm_2021}, \citealt{Hirano_2021}; G 9-40 b: \citealt{Stefansson_2020}, \citealt{Luque_2022b}). Here, we present the confirmation of a new keystone planet: TOI-1695 b. Our analysis presented herein includes a mass constraint from observations from the High Accuracy Radial Velocity Planet Searcher - North (HARPS-N) spectrograph. Paired with a radius constraint from the TESS data, we are able to measure the planet's bulk composition and place constraints on the emergence mechanism of the radius valley around early M dwarfs.

In Section~\ref{Observations} we present the TESS light curve and our suite of follow-up observations. In Section~\ref{stellar_section} we present the properties of the host star TOI-1695. In Section~\ref{Data Analysis and Results} we present our data analysis and results. In Section~\ref{Discussion} we discuss the importance of our findings in the context of the greater science questions and in Section~\ref{Summary} we conclude with a summary.

\section{Observations} \label{Observations}

\begin{figure*}
    \centering
    \includegraphics[width=\textwidth]{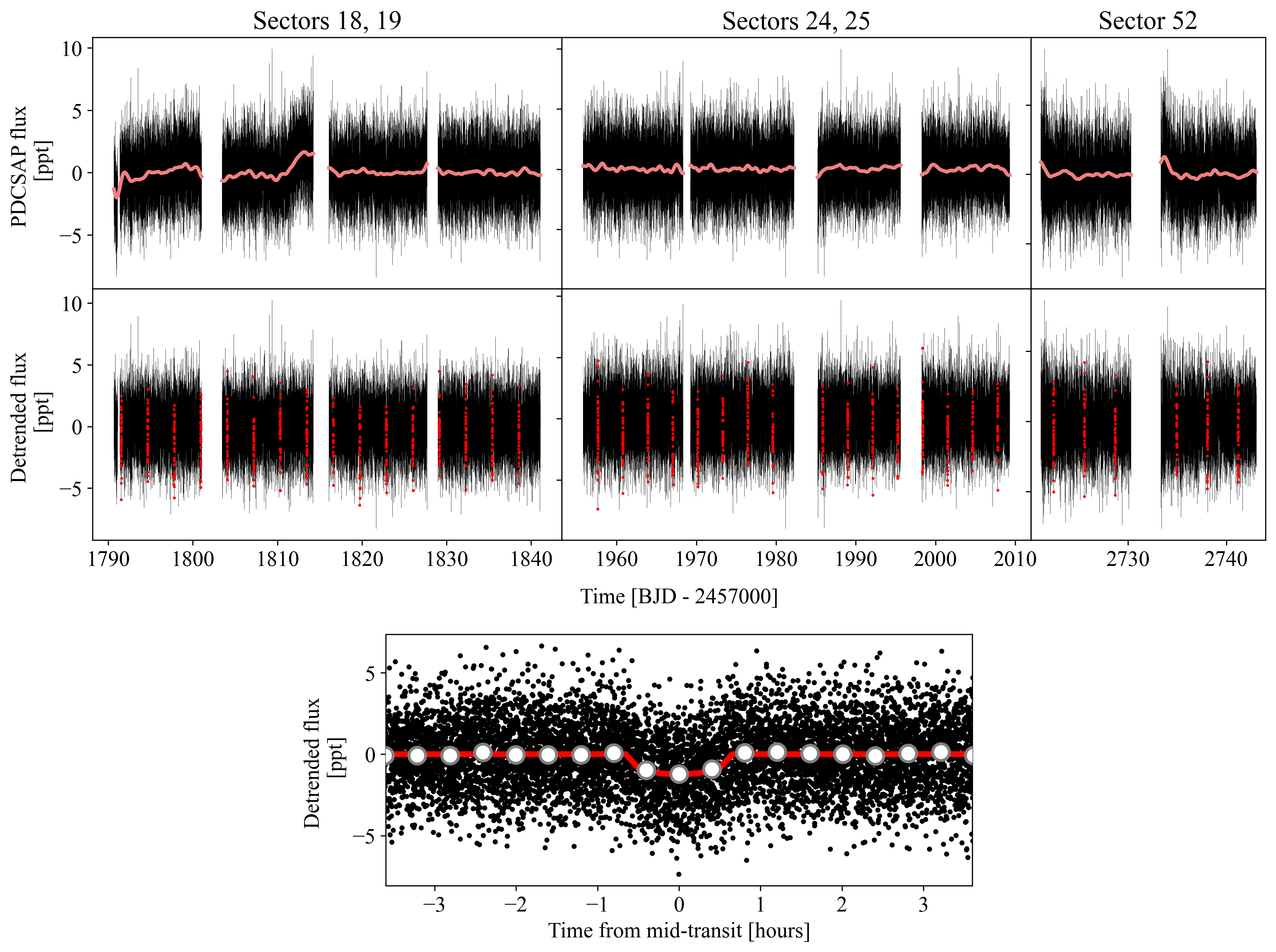}
    \caption{TESS {\tt\string PDCSAP} light curve of TOI-1695 from Sectors 18, 19, 24, 25, and 52. $\mathbf{Top\ row}$: the dilution-, systematic-, and background-corrected {\tt\string PDCSAP} light curve overlaid with our mean GP model of the residual correlated noise (pink curve). $\mathbf{Middle\ row}$: The {\tt\string PDCSAP} light curve detrended after subtraction of our mean GP model. In-transit measurements are highlighted in red. $\mathbf{Bottom\ panel}$: the phase-folded transit light curve of TOI-1695 b with 24-minute bins. The maximum a-posteriori transit model is overlaid in red and the white markers depict the binned light curve.}
    \label{fig:TESS_curves}
\end{figure*}

\subsection{TESS Photometry} \label{TESS Photometry}
TESS is an ongoing NASA mission to survey the entire sky to search for nearby transiting planets \citep{Ricker_2015}. The spacecraft orbits the Earth in an elliptical, 2:1 linear synchronous orbit with a period of 13.7 days. Annual observation cycles are split into sectors lasting two orbits, which is about 27 days. The detector consists of four contiguous CCD cameras, each covering a 24$^o$ x 24$^o$ field of view, making up a 24$^o$ by 96$^o$ strip aligned along ecliptic latitude lines. TESS uses the MIT Lincoln Laboratory CCID-80 detector with a depletion depth of 100 $\mu$m, allowing for sensitivity just redward of 1000 nm. At the blue end, the spectral response is limited by a longpass filter with a cut-on wavelength of 600 nm \citep{Sullivan_2015}. In years 1 and 2, the data were stored with a short cadence of 2 minutes and long cadence of 30 minutes. In year 3, an additional ultra-short cadence of 20 s is avaliable, and the long cadence was reduced to 10 minutes. 

TOI-1695 was observed in five nonconsecutive TESS sectors between UT 2019 November 2 and UT 2022 June 13 at two-minute cadence. The full baseline of the TESS observations is 952 days. TESS observations occurred in CCD 2 on camera 3 in Sector 18 (UT 2019 November 2-November 27), in CCD 1 on camera 3 in sector 19 (UT 2019 November 27-December 24), in CCD 4 on camera 4 in sector 24 (UT 2020 April 16-May 13), in CCD 3 on camera 4 in sector 25 (UT 2020 May 13-June 8), and in CCD 3 on camera 4 in sector 52 (UT 2022 May 18-June 13).

The TESS images were processed by the NASA Ames Science Processing Operations Center \citep[SPOC;][]{Jenkins_2016}, which produced the Simple Aperture Photometry \citep[{\tt\string SAP};][]{Twicken_2010, Morris_2020} and Presearch Data Conditioning Simple Aperture Photometry ({\tt\string PDCSAP}) light curves per sector. The latter were corrected for systematic uncertainties exhibited by all sources within the field \citep{Smith_2012, Stumpe_2012, Stumpe_2014}. The light curves are corrected for dilution during the SPOC processing with TOI-1695 suffering low levels of contamination as indicated by its average dilution correction factor across all five sectors of 0.983. We only consider reliable TESS measurements for which the measurement’s quality flag is equal to zero. TOI-1695's {\tt\string PDCSAP} light curve is depicted in Figure \ref{fig:TESS_curves} and shows no compelling signs of coherent photometric variability, in particular from rotation.

Following {\tt\string PDCSAP} light curve construction, the SPOC conducted a transit search using the Transiting Planet Search (TPS) module \citep{Jenkins_2002, Jenkins_2010}. A repeating transit-like signal with a reported period of 3.13 days was detected in all five sectors independently. A total of 38 transit events were observed over the five TESS sectors and are highlighted in Figure \ref{fig:TESS_curves}. The signal passed a set of internal data validation tests \citep{Twicken_2018} and was fit with a preliminary limb-darkened transit model. After a review of the diagnostic tests in the data validation reports, the TESS Science Office classified the planet candidate as TOI-1695.01 \citep{Guerrero_2021}.  The SPOC reported a preliminary $R_p/R_s$ value of 0.0316, which corresponds to a planetary radius of 1.78 $R_\oplus$ using our adopted stellar radius of $0.515\ R_\odot$ (see Section~\ref{stellar_section}).

\subsection{ASAS-SN Photometry} \label{sect:asassn}
Active regions on the surface of a rotating star will introduce a time-varying signal in both photometric and in precise radial velocity measurements. Our ability to construct complete models of each of these datasets therefore benefits from a-priori knowledge of the stellar rotation period \Prot{.} Because the TESS light curve does not exhibit any signature of stellar rotation (Figure \ref{fig:TESS_curves}), we queried the All-Sky Automated Survey for Supernovae \citep[ASAS-SN;][]{Shappee_2014,Kochanek_2017} data archive to search for long-term photometic monitoring of TOI-1695. ASAS-SN is a global network of 24 telescopes, hosted by the Las Cumbres Observatory, whose ongoing goal is to monitor the entire sky on a continuous basis to search for transient phenomena. 

\begin{figure}
    \centering
    \includegraphics[width=\hsize]{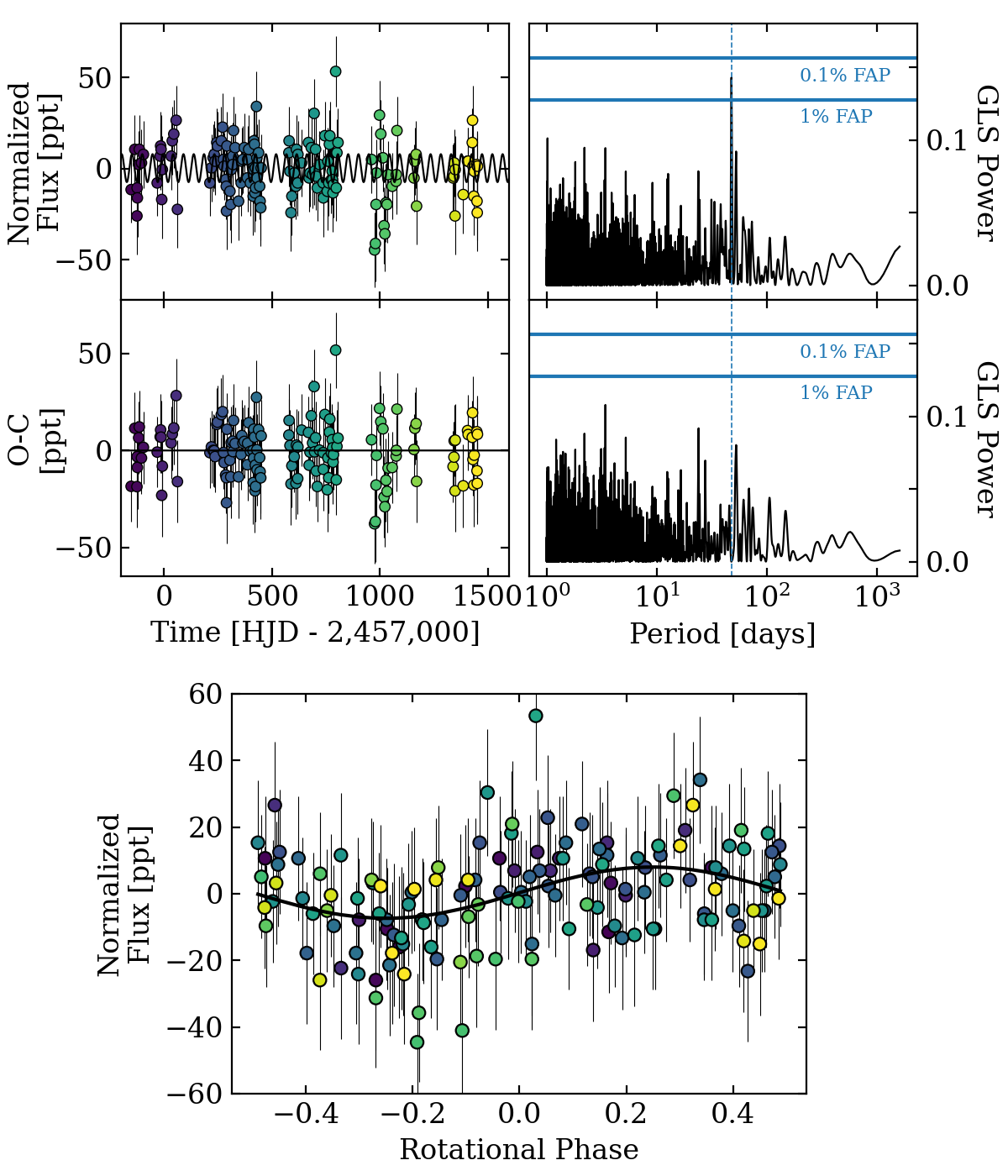}
    \caption{ASAS-SN photometric monitoring of TOI-1695. $\mathbf{Top\ row:}$ the full $V$-band photometric light curve over its four-year baseline, overplotted with our best-fit sinusoidal model of stellar rotation with an amplitude of 7.7 ppt. The marker colors indicate the epoch of each observation. The right panel depicts the GLS periodogram of the light curve and reveals a strong signal at 48 days. $\mathbf{Middle\ row:}$ the residual light curve after the removal of the best-fit stellar rotation model. The corresponding GLS periodogram reveals no significant residual periodicities. $\mathbf{Bottom\ panel:}$ the light curve and rotation model phase-folded to the measured stellar rotation of 47.7 days.}
    \label{fig:asassn}
\end{figure}

Our data archive search revealed that TOI-1695 was monitored throughout the ASAS-SN campaign in the $V$-band for more than four years from UT 2014 July 8 to UT 2018 November 29 (Figure~\ref{fig:asassn}). The nightly cadence of the light curve is sufficient to detect rotational variability on timescales that exceed a few days, such as what we expect for TOI-1695 given the lack of photometric variability in its TESS light curve. We computed the generalized Lomb-Scargle periodogram \citep[GLS;][]{Zechmeister_2009} of these data and uncover a strong periodicity at approximately 48 days that is not seen in the light curve's window function (not shown in Figure~\ref{fig:asassn}). We interpret this periodic signal as the stellar rotation period and fit the ASAS-SN photometry with a sinusoidal function as shown in the bottom row of Figure~\ref{fig:asassn}. From this fit we measure a photometric amplitude of 7.7 ppt and \Prot{} $=47.7\pm 2.2$ days. We note that the GLS of the light curve residuals and does not show any signature of a significant residual periodicity. We also note that this value of \Prot{} is consistent with the expectation from population studies of inactive early-M dwarf rotation periods \citep[e.g.][]{Newton_2016}.
We will use this measurement of \Prot{} as a prior in our forthcoming data analyses.

\subsection{Reconnaissance Spectroscopy} \label{recon_spec}
We obtained a pair of reconnaissance spectra of TOI-1695 on UT 2020 February 2 and 13 using the Tillinghast Reflector \'Echelle Spectrograph \citep[TRES;][]{tres}. These observations were coordinated as part of the TESS Follow-up Observing Program (TFOP). TRES is an $R=44,000$ fibre-fed optical spectrograph (310-910 nm) mounted on the 1.5m Tillinghast Telescope at the Fred Lawrence Whipple Observatory on Mount Hopkins, Arizona. The spectra were reduced and extracted following the standard procedure \citep{Buchhave_2010} and were subsequently cross-correlated against a custom template of Barnard's star over a range of \vsini\ values \citep{Winters_2018}. 
These spectra reveal that TOI-1695 is single-lined, lacks a measurable signal from rotational broadening (i.e. \vsini{} $< 3.4$ km s$^{-1}$), and exhibits $H\alpha$ in absorption. Taken together, these findings are consistent with the absence of strong photometric variability and confirm that TOI-1695 is likely a slowly-rotating and chromospherically inactive star.

We also measured stellar radial velocities (RV) at opposing quadrature phases of $-59.766\pm 0.081$ km s$^{-1}$ and $-59.892\pm 0.056$ km s$^{-1}$, indicating that there is no significant RV variation that would have been produced if the system were a spectroscopic binary. As such, the transit-like signal TOI-1695.01 remains a viable planet candidate that we continue to vet observationally in the following subsections. 

\subsection{Ground-Based Photometry} \label{sect:sg1}

\begin{figure}
    \centering
    \includegraphics[width=0.9\hsize]{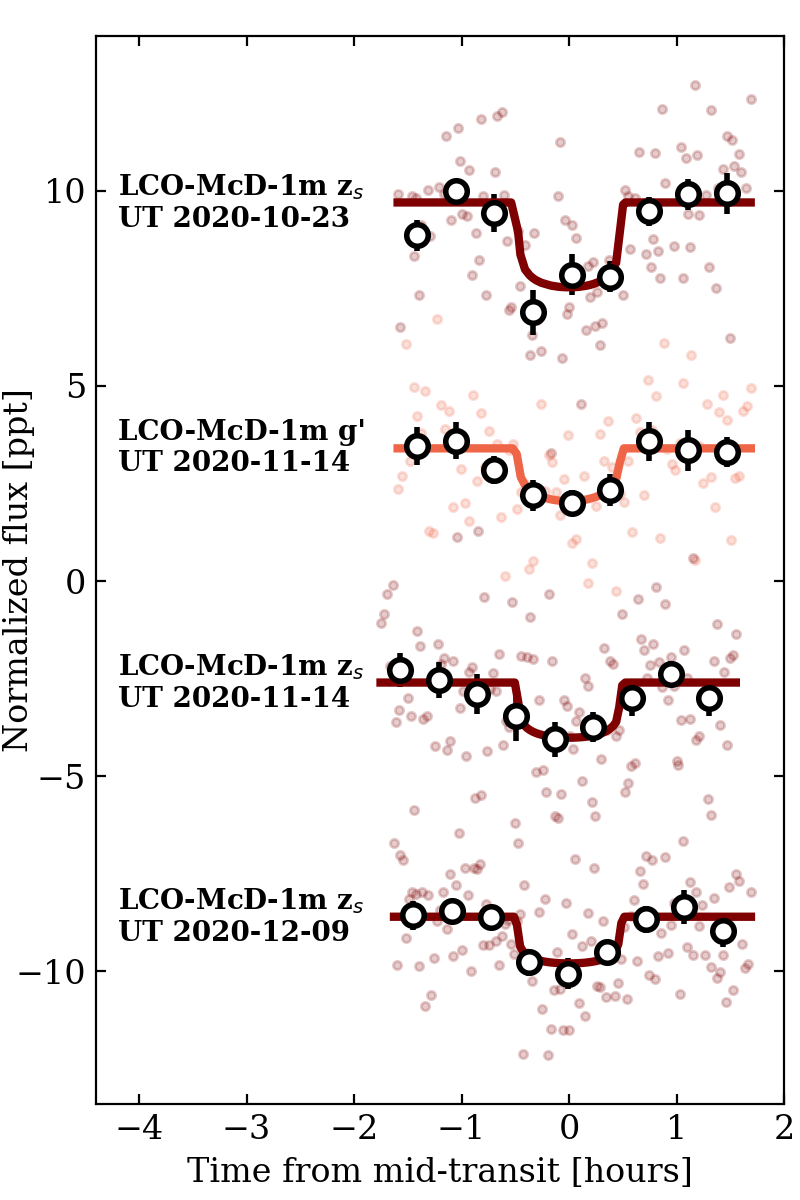}
    \caption{Ground-based transit light curves of TOI-1695.01 taken as part of TFOP. The solid curves depict the optimized transit model fits with all model parameters fixed other than the baseline flux, mid-transit time, and planet-to-star radius ratio. Annotated next to each light curve is the telescope facility, passband of the observation, and UT observation date. Here we only depict the four light curves that were confirmed to have temporal coverage over a transit event.}
    \label{fig:sg1}
\end{figure}

With a pixel scale of $21\arcsec$ pixel$^{-1}$, and photometric apertures that typically extend out to roughly $1\arcmin$, TESS commonly produces light curves with multiple blended sources. Indeed, there are 68 sources from Gaia DR3 within 2\farcm5 of TOI-1695, with TOI-1695 being the brightest source ($\Delta G = 2.693$) and requiring only a marginal dilution correction in the TESS bandpass (i.e. 0.983). To resolve this, we acquired seeing-limited ground-based transit follow-up photometry of TOI-1695.01. These observations were taken as part of the TFOP Sub Group 1 \citep[TFOP SG1;][]{collins:2019}\footnote{https://tess.mit.edu/followup} to rule out or identify nearby eclipsing binaries (NEBs) and to check for the transit-like event on-target using the greater spatial resolution compared to the TESS images. Our data also constrain the chromatic transit depth across complementary optical filter bands. We used the {\tt TESS Transit Finder} to schedule our transit observations and the photometric data were extracted using {\tt AstroImageJ} \citep{Collins:2017}.

\subsubsection{MLO}
We observed a predicted full transit window of TOI-1695.01, according to the initial SPOC TESS sector 19 ephemeris, in $I$-band on UT 2020 August 21 from the Maury Lewin Observatory (MLO) 0.36\,m telescope near Glendora, CA. The telescope is equipped with a $3326\times2504$ SBIG STF8300M camera having an image scale of $0\farcs84$ per pixel, resulting in a $23\arcmin\times17\arcmin$ field of view. The images were calibrated and the photometric data were extracted using {\tt AstroImageJ}. The data were not sensitive at the level of the expected shallow event on TOI-1695, but we searched the nearby field within $2\farcm5$ and did not detect an obvious NEB that might be causing the TESS detected event. However, after the availability of a more precise ephemeris from a joint SPOC analysis that included subsequent TESS sectors 18, 19, 24, and 25, it was determined that these observations did not cover a transit window.

\subsubsection{LCOGT}
We observed four predicted transit windows of TOI-1695.01 from the Las Cumbres Observatory Global Telescope \citep[LCOGT;][]{Brown:2013} 1.0\,m network node at McDonald Observatory in Texas, USA, on UT 2020 August 24, 2020 October 23, 2020 November 14, and 2020 December 9. All but one observation was conducted in Pan-STARRS $z_s$-band, with one of two observations on UT 2020 November 14 being taken in the Sloan $g'$-band. The 1\,m telescopes are equipped with $4096\times4096$ SINISTRO cameras having an image scale of $0\farcs389$ per pixel, resulting in a $26\arcmin\times26\arcmin$ field of view. The images were calibrated by the standard LCOGT {\tt BANZAI} pipeline \citep{McCully:2018}.

The first LCOGT observation on UT 2020 August 24 should have been sensitive to the 1.2 ppt event if it occurred on TOI-1695, but the data likely ruled out an event on or off target that would have been deep enough to cause the event detected by the SPOC pipeline. However, like the MLO observations one predicted orbit earlier, these observations turned out to be out-of-transit relative to the later multi-sector SPOC ephemeris. The four remaining LCOGT observations were conducted according to the precise multi-sector SPOC ephemeris and achieved continous coverage across the full transit events. We used {\tt AstroImageJ} to extract the photometric data using circular photometric apertures with radii in the range $4\farcs3$ to $5\farcs8$. All of the TOI-1695 apertures exclude flux from the nearest known Gaia DR3 and TESS Input Catalog neighbor (TIC 629325854) $14\farcs6$ East. We perform a least squares fit to the individual light curves using a combined transit plus systematics model. We construct the latter using a linear combination of time, the FWHM of the PSF, and the sky background. 

We detect the transit event within the TOI-1695 photometric apertures in the three $z_s$-band and one $g'$-band light curves (see Figure~\ref{fig:sg1}). We find that in all but the $z_s$ light curve taken on UT 2020 October 23, the measured values of $R_p/R_s \in [0.033,0.036]$ are consistent with the value measured from TESS. The one exceptional light curve produced an anomalously large $R_p/R_s=0.045$, which we consider to be an outlier and not a true chromatic effect due to its inconsistency with the other $z_s$-band light curves. Our results confirm that the transit event of TOI-1695.01 occurs on target such that we are able to rule out NEBs and continue to interpret TOI-1695.01 as a viable planet candidate. 


We note that we do not include these observations in our global transit analysis (Section~\ref{TESS Transit Analysis}) because of strong dependence of each light curve on the exact systematics model used. Plus, because of the multi-year TESS baseline, which extends well beyond our most recent ground-based observation, the ground-based light curves presented herein do not provide stronger constraints on the planet candidate's orbital period when compared to TESS alone. 

\subsection{High-Resolution Imaging} \label{hires_imaging}

\begin{figure}
    \centering
    \includegraphics[width=\hsize]{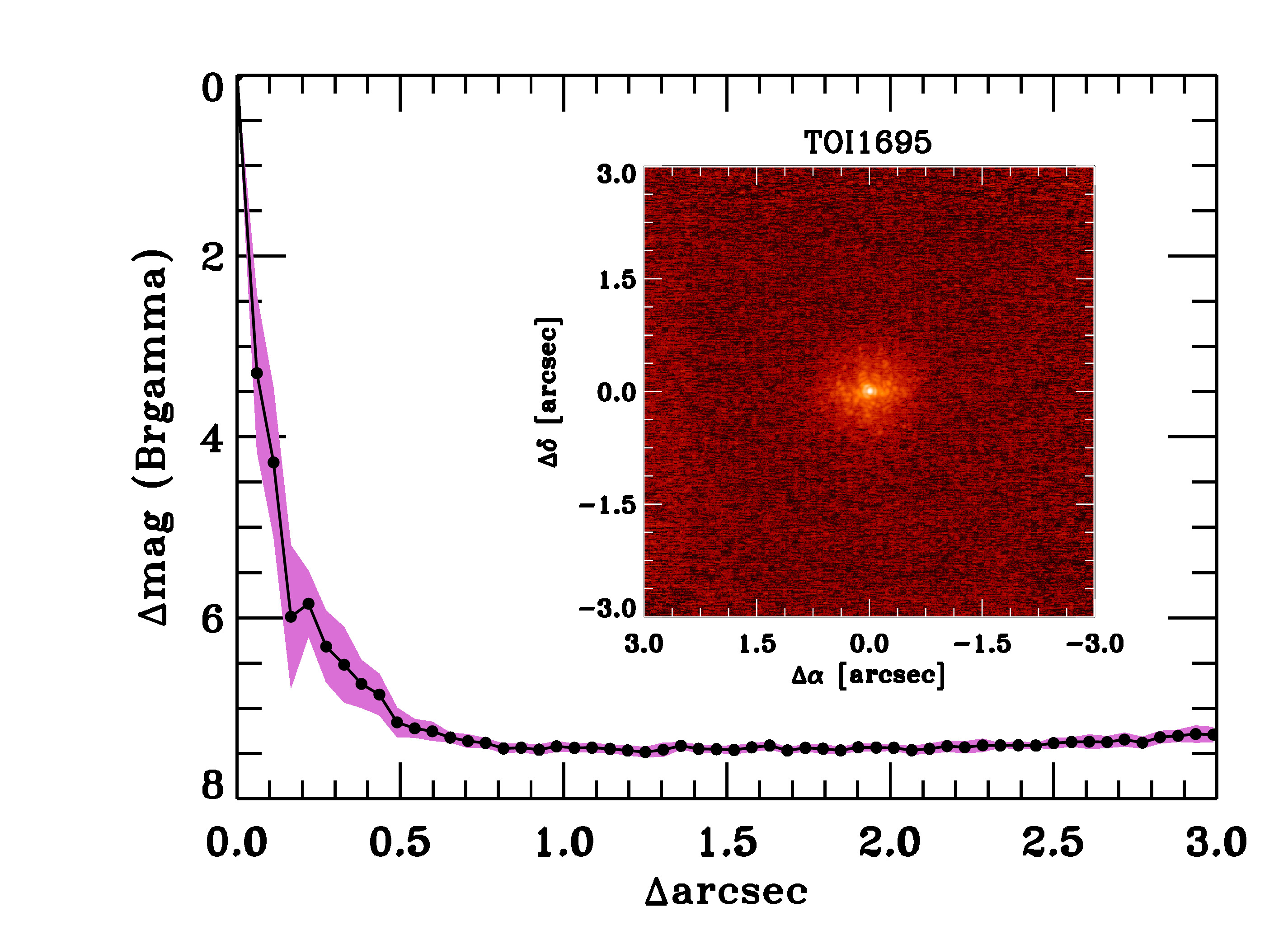}
    \includegraphics[width=\hsize]{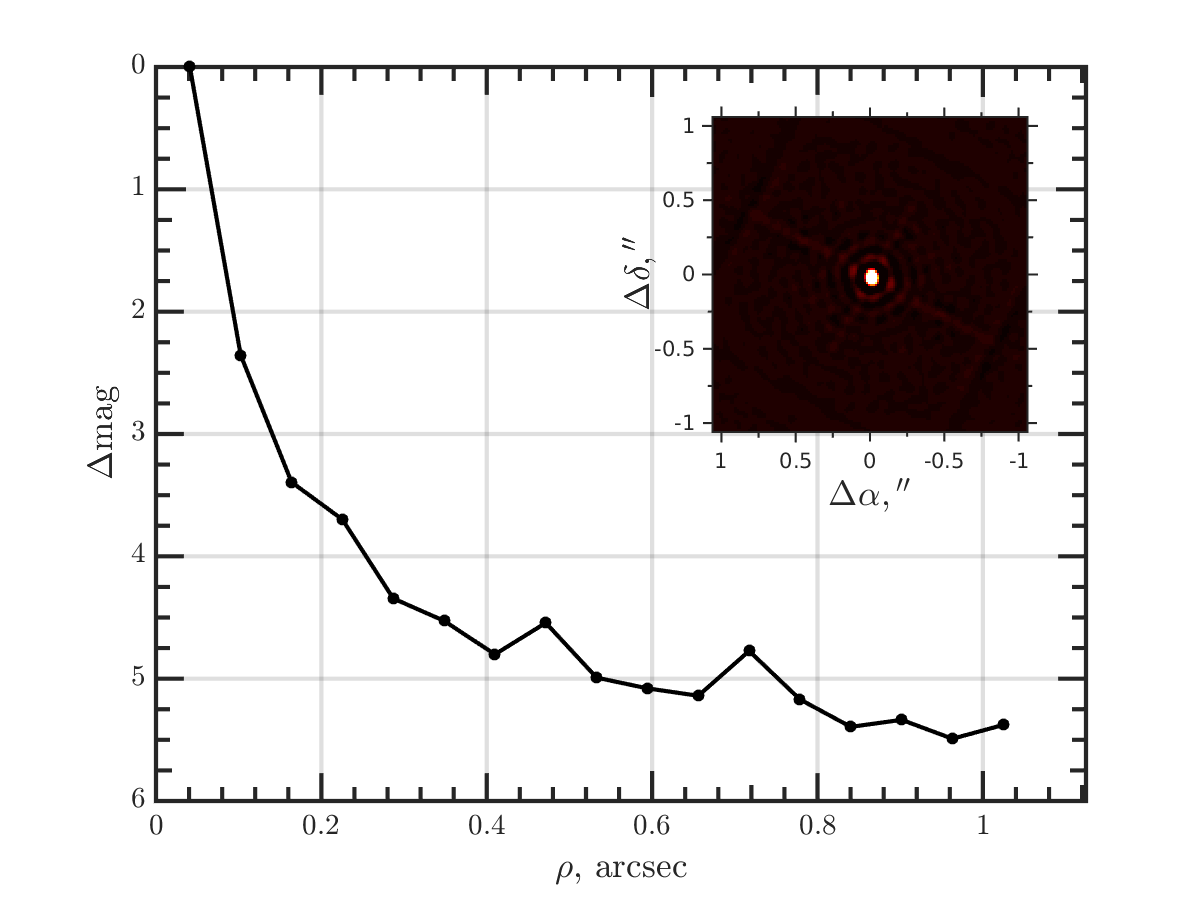}
    \caption{Companion sensitivity from high-resolution imaging. Top: the results from Keck/NIRC2 adaptive optics imaging in the $Br-\gamma$ filter. The black points represent the 5$\sigma$ limits and are separated in steps of 1 FWHM ($\sim 0\farcs054$); the purple represents the azimuthal dispersion (1$\sigma$) of the contrast determinations. The inset image is of the primary target showing no additional companions to within 3\arcsec\ of the target. Bottom: the results from SAI speckle polarimetry in the $I_c$-band. The black curve depicts the $5\sigma$ contrast limits and the inset shows the combined intensity image on a linear scale.}
    \label{fig:sg3}
\end{figure}

\subsubsection{Keck/NIRC2}
Following \cite{ciardi2015}, we assessed the possible contamination of the TESS light curve by bound or unbound companions using high-resolution adaptive optics (AO) imaging from NIRC2 on Keck II \citep{wizinowich2000}. We observed TOI-1695 on UT 2020 May 28 in the narrow-band $Br-\gamma$ filter with an integration time of four seconds with one coadd per frame for a total of 36 seconds on target. Our AO data were processed and analyzed following the standard procedure \citep{furlan2017}, which includes the calculation of the $5\sigma$ contrast curve via source injection (see Figure \ref{fig:sg3}).

We detect no additional companions around TOI-1695 given the sensitivity of our data. We demonstrate sensitivities down to $\sim 3.5$ mag at $0\farcs06$ (2.7 au) and $\sim 7$ mag at $0\farcs5$ (22 au). These contrast limits indicate that there are likely no stellar companions down to M6-L9 spectral types within $3\farcs0$ of the position of TOI-1695. From these results we conclude that TOI-1695.01 remains a viable planet candidate and is worthy of the time investment to obtain precise RV measurements for planet confirmation and planetary mass measurement.

\subsubsection{SAI Speckle Polarimetry}
We observed TOI-1695 on 2020 October 29 UT with the Speckle Polarimeter \citep{Safonov_2017} on the 2.5~m telescope at the Caucasian Observatory of Sternberg Astronomical Institute (SAI) of Lomonosov Moscow State University. SPP uses Electron Multiplying CCD Andor iXon 897 as a detector. The atmospheric dispersion compensator allowed observation of this relatively faint target through the wide-band $I_c$ filter. The power spectrum was estimated from 4000 frames with 30 ms exposure. The detector has a pixel scale of 20.6 mas pixel$^{-1}$ and the angular resolution was 89 mas. Consistent with our results from Keck/NIRC2, we do not detect any stellar companions brighter than $\Delta I_c=4$ and $5.4$ at separations of $0\farcs25$ and $1\farcs0$, respectively.

\subsection{Precise Radial Velocity Measurements} \label{sect:harpsn}
We obtained 49 spectra of TOI-1695 with the HARPS-N optical \'echelle spectrograph at the 3.6m Telescopio Nazionale Galileo on La Palma in the Canary Islands. HARPS-N has a resolving power of $R$ = 115,000 and is stabilized in pressure and temperature, thus enabling sub-meter per second instrumental stability \citep{Consentino_2012}. We observed TOI-1695 over 428 days between UT 2020 December 6 and UT 2022 February 7 as part of the HARPS-N Guaranteed Time Observations program. We fixed the exposure time to 1800 s throughout the campaign, which yielded a median signal-to-noise ratio (S/N) per order of 15.9 across all orders redward of aperture 18 (440-687 nm).

We reduced the spectra using version 3.7 of the HARPS-N Data Reduction Software \citep[DRS;][]{Lovis_2007}, which includes an automated RV extraction using an M0 template in the cross-correlation function (CCF). We opted to conduct a separate RV extraction using the template-matching algorithm \texttt{TERRA} \citep{Anglada_2012}, which has been shown to outperform the CCF method on M dwarfs \citep[e.g.][]{Anglada_2012,Astudillo_2015}.  \texttt{TERRA} constructs an empirical master spectral template by coadding the individual HARPS-N spectra after being translated into the barycentric frame. We ignore spectral regions in which the atmospheric transmission is $<99$\% and only consider \'echelle orders redward of aperture 18, following the recommended procedure for M dwarfs \citep{Anglada_2012}. From the remaining spectral regions, we compute the RV shift of each spectrum via least-squares matching to the master template. We obtain a median RV uncertainty and raw RV rms of 1.81 m s$^{-1}$ and 6.32 m s$^{-1}$, compared to values of 6.11 m s$^{-1}$ and 8.02 m s$^{-1}$, respectively, from the DRS.

We also measure logR'$_{HK}$ = -4.74 $\pm$ 0.41 from our master template. From this we derive an expected \Prot{} of 27.8$^{+26.8}_{-13.2}$ days from the activity-rotation relation from \cite{Astudillo_2017}, which is consistent with the measured \Prot{} from ASAS-SN. TOI-1695 is in the unsaturated chromospheric activity regime, which is consistent with an absence of observed broadening in the HARPS-N spectrum (\vsini\ $<1.3$ km s$^{-1}$) and with its H$\alpha$ being seen in absorption.

\begin{figure*}
    \centering
    \includegraphics[width=\textwidth]{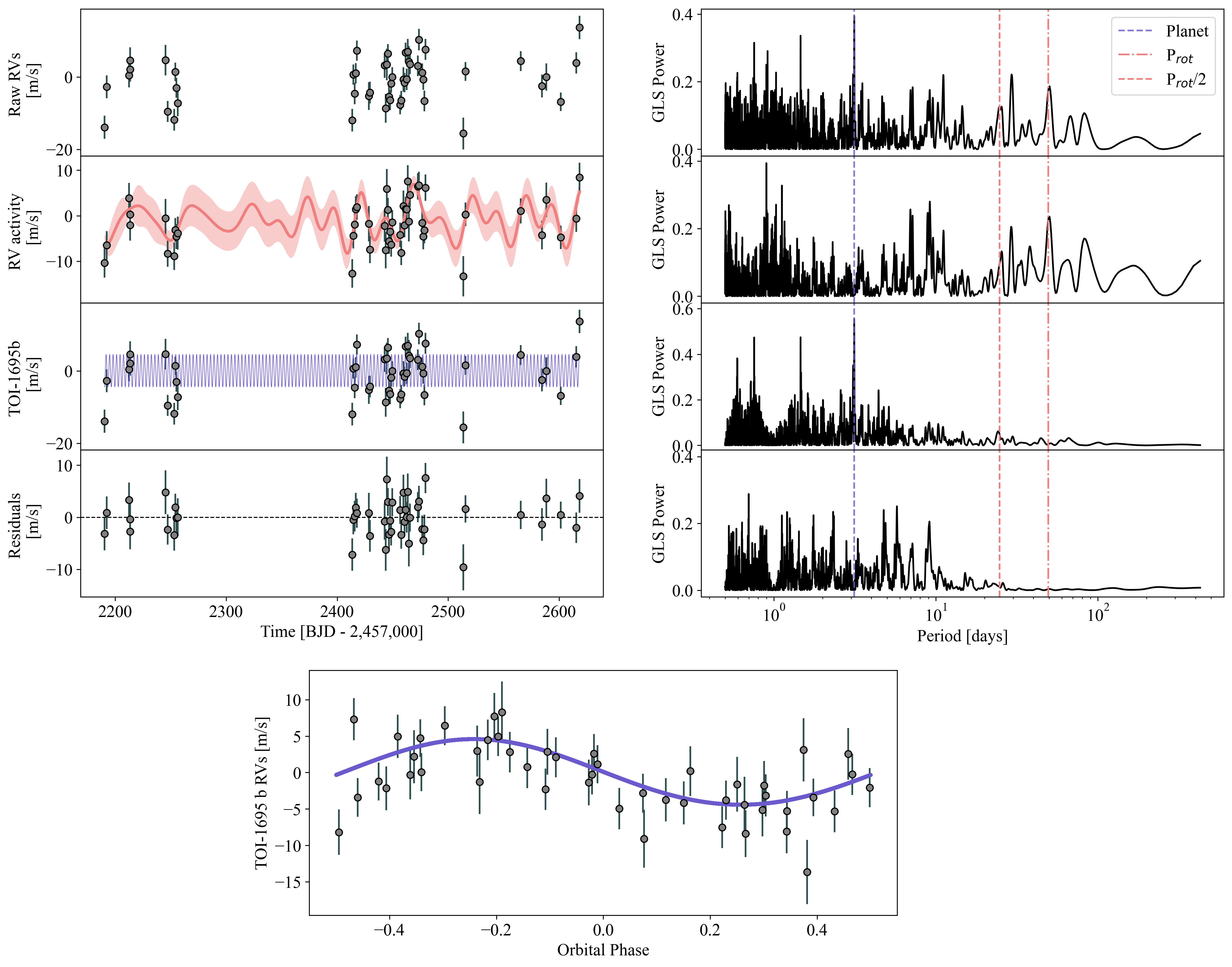}
    \caption{The TOI-1695 HARPS-N RVs and model components along with their corresponding GLS periodogram. $\mathbf{Top\ row}$: the raw RVs. The vertical blue dashed line in the periodogram highlights the orbital period of TOI-1695 b while the red dashed lines highlight the stellar rotation period and its first harmonic. $\mathbf{Second\ row}$: the activity component of our RV model (i.e. raw RVs minus the maximum a-posteriori Keplerian model) overlaid with the mean GP model (red curve). The shaded region represents the standard deviation on the GP. $\mathbf{Third\ row}$: the RV signal from TOI-1695 b overlaid with the maximum a-posteriori Keplerian solution for TOI-1695 b (blue curve). $\mathbf{Fourth\ row}$: the RV residuals. $\mathbf{Bottom\ panel}$: the activity-corrected RVs phase-folded to the orbital period of TOI-1695 b. The RV measurement uncertainties throughout include the contribution from the additive scalar jitter term $s_{RV}$.}
    \label{fig:RVs}
\end{figure*}

The raw RVs from \texttt{TERRA} are shown in the upper row of Figure~\ref{fig:RVs}. The corresponding GLS periodogram of the raw RVs clearly exhibits a strong periodicity at the 3.13-day period of the planet candidate TOI-1695.01. We therefore proceed by interpreting TOI-1695.01 as a validated planet, which we will now refer to as TOI-1695 b for the remainder of this paper.

\section{Stellar Characterization} \label{stellar_section}

TOI-1695 (TIC 422756130) is an early M dwarf located in the northern sky at a distance of 44.993 $\pm$ 0.028 pc \citep{Bailer-Jones_2021}. The star has no known binary companions and no comoving sources in Gaia DR3 \citep[see Sections \ref{recon_spec} to \ref{hires_imaging};][]{Gaia_DR3}. Gaia DR3 also reports astrometric excess noise of 88 $\mu$as, a RUWE statistic of 1.08, and a null non-single star flag indicating no clear departure from a single-star model for this source. 

\begin{figure}
    \centering
    \includegraphics[width=\hsize]{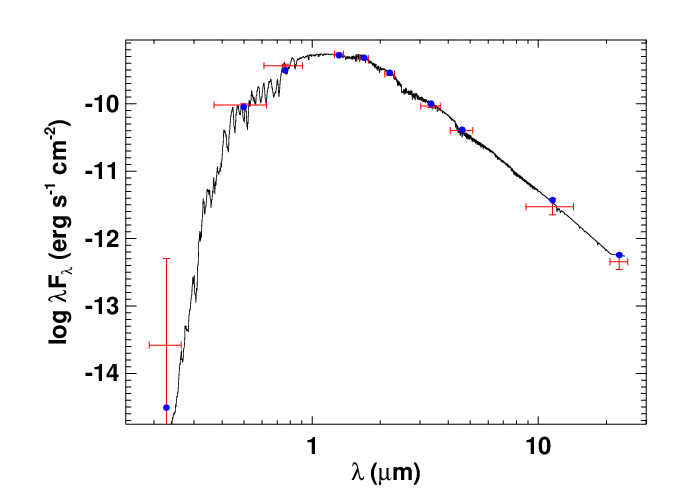}
    \caption{The spectral energy distribution and best-fit stellar atmosphere model of \name{.} The red markers
    depict the photometric measurements with horizontal error bars that depict the effective width of each passband. The black curve depicts the best-fit stellar atmosphere model with \teff{} $=3630$ K. The blue circles depict the model fluxes integrated over each passband.}
    \label{fig:SED}
\end{figure}

We performed a preliminary analysis of the broadband spectral energy distribution (SED) of the star together with the Gaia DR3 parallax \citep[with no systematic offset applied; see e.g.,][]{StassunTorres:2021} following the procedures described in \citet{Stassun:2016,Stassun:2017,Stassun:2018}. We obtained the $JHK_s$ magnitudes from 2MASS, the W1-W4 magnitudes from WISE, the $G_{\rm BP},\ G,\ G_{\rm RP}$ magnitudes from Gaia, and the NUV flux from GALEX. Together, the available photometry spans the stellar SED over the wavelength range 0.2--22~$\mu$m (Figure~\ref{fig:SED}).  

We performed a fit using NextGen stellar atmosphere models \citep{Hauschildt_1999}, with the effective temperature (\teff{)} and metallicity ([Fe/H]) as free parameters (the surface gravity, $\log g$, has very little influence on the broadband SED). We limited the extinction $A_V$ to the full line-of-sight value from the Galactic dust maps of \citet{Schlegel:1998}. The resulting fit has a reduced $\chi^2$ of 1.3, with best-fit \teff{} $= 3630 \pm 50$~K and [Fe/H] = $0.0 \pm 0.5$. Integrating the model SED gives the bolometric flux at Earth of $F_{\rm bol} = 6.82 \pm 0.24 \times 10^{-10}$ erg~s$^{-1}$~cm$^{-2}$. Taking the $F_{\rm bol}$ together with the Gaia parallax directly gives the luminosity, $L_{\rm bol} = 0.0431 \pm 0.0015\ L_\odot$. Similarly, the $F_{\rm bol}$ together with the \teff{} and the parallax gives the stellar radius $R_s = 0.525 \pm 0.017\ R_\odot$. This value is consistent with the stellar radius from the empirically derived $K_s$-band radius-luminosity relation from \cite{Mann_2015} ($0.515\pm 0.015$). Similarly, we find a consistent \teff\ $=3690\pm 50$ K using the color-\teff\ relation from \cite{Mann_2015}, which we evaluate using the $G_{\rm BP}-G_{\rm RP}$ color and adopt as our final value. Finally, we derived a stellar mass of $M_s = 0.513 \pm 0.012\ M_\odot$ using the empirical $K$-band mass-luminosity relation from \cite{Mann_2019}.

The astrometric, photometric, and physical stellar parameters are reported in Table \ref{tab:stellar_params}.

\input{stellar_table}

\section{Data Analysis and Results} \label{Data Analysis and Results}

\input{point_estimates_table}

We now seek to measure the fundamental orbital and physical planetary parameters of TOI-1695 b by first fitting a Gaussian process (GP) plus transit model to the TESS light curve, followed by a separate RV analysis with data from HARPS-N. The planet parameter posteriors from the transit analysis were used as priors for the RV analysis. Priors for the planet model parameters and GP hyperparameters are presented in Table \ref{tab:priors}.

\input{priors_table}

\subsection{TESS Transit Analysis} \label{TESS Transit Analysis}
We first model the raw TESS {\tt\string PDCSAP} light curve (Figure \ref{fig:TESS_curves}, top row) in which the planet candidate TOI-1695.01 was originally detected. The {\tt\string PDCSAP} light curve has already undergone systematics corrections via a linear combination of cotrending basis vectors; however, some low-amplitude and temporally-correlated signals that are unrelated to planetary transits persist. We model this residual systematic noise in the {\tt\string PDCSAP} curve using a GP simultaneously with our transit model. We employ the {\tt\string exoplanet} software package \citep{ForemanMackey_2019} to sample the posterior of the joint GP and transit model parameters at each step in our Markov Chain Monte Carlo (MCMC) simulation. The {\tt\string exoplanet} package uses the {\tt\string STARRY} package \citep{Luger_2019} to compute analytical transit models and {\tt\string celerite} \citep{ForemanMackey_2017} to evaluate the likelihood of the GP model.

We adopt a covariance kernel of the form of a stochastically-driven, damped, simple harmonic oscillator in Fourier space. The power spectral density of the kernel is

\begin{equation}
\label{gp kernel}
    S(\omega) = \sqrt{\frac{2}{\pi}} \frac{S_0 \omega_0^4}{(\omega^2 - \omega_0^2)^2 + \omega_0^2 \omega^2/Q^2}.
\end{equation}

\noindent The kernel is parameterized by the undamped period of the oscillator $\rho = 2\pi/\omega_0$ where $\omega_0$ is the undamped angular frequency ($\omega$ is the angular frequency); the standard deviation of the process $\sigma_{TESS} = \sqrt{S_0 \omega_0 Q}$; and the fixed quality factor Q = 1/$\sqrt{2}$. Our GP model is jointly fit with a transit model for TOI-1695 b with the following free parameters: stellar mass $M_s$, stellar radius $R_s$, a reparameterization of the quadratic limb-darkening coefficients $q$1 and $q$2 \citep{Kipping_2013}, orbital period $P$, time of mid-transit $t_0$, log transit depth ln$\delta$, baseline flux $f_{0, TESS}$, impact parameter $b$, eccentricity $e$, and argument of periastron $\omega$. $e$ is reparameterized with two shape parameters and sampled from a beta distribution $\mathcal{B}$ as described in \cite{Kipping_2013b}. Our full TESS transit model therefore contains the following 13 parameters: \{ln$\rho$, ln$\sigma_{TESS}$, $M_s$, $R_s$, $q$1, $q$2, ln$P$, $t_0$, ln$\delta$, $f_{0, TESS}$, $b$, $e$, $\omega$\}.

We execute an MCMC to sample the joint posterior probability density function (PDF) of our full set of model parameters using the {\tt\string PyMC3} MCMC package \citep{Salvatier_2016} within {\tt\string exoplanet}. The MCMC is initialized with two simultaneous chains, each with 1500 tuning steps and 1000 draws in the final sample. Point estimates of the maximum a-posteriori values from the marginalized posterior PDFs of the GP hyperparameters are selected to construct the GP predictive distribution, whose mean function we adopt as our detrending model of the {\tt\string PDCSAP} light curve. This mean detrending function and the detrended light curve are both shown in Figure \ref{fig:TESS_curves}. Similarly, we recover the maximum a-posteriori point estimates of the transit model parameters to construct the transit model shown in the bottom panel of Figure \ref{fig:TESS_curves}. Median maximum a-posteriori values and uncertainty point estimates from the 16$^{\rm th}$ and 84$^{\rm th}$ percentiles for all model parameters are reported in Table \ref{tab:point_estimates}. Recall that we do not consider ground-based transits jointly with our TESS transit analysis because they are susceptible to residual systematic uncertainties and our longest transit baseline is spanned by TESS sectors 18 through 52.

\subsection{RV Analysis} \label{RV Analysis}
We impose strong priors on $P$ and $T_0$ derived from our TESS analysis. The raw RVs and their GLS periodograms are shown in Figure \ref{fig:RVs}. The periodic signal induced by the orbit of TOI-1695 b is clearly visible at 3.13 days. The rotation period \Prot{} = 48 days of TOI-1695 is well constrained from 4 years of ASAS-SN data (see Figure \ref{fig:asassn}), and a moderate signal is observed near this period in the GLS periodogram, consistent with the star being inactive. We simultaneously fit the observed data with a Keplerian orbit and a quasi-periodic GP regression model of stellar activity. We adopt a GP covariance kernel of the form of two stochastically-driven, damped, simple harmonic oscillators in Fourier space, both described by Equation \ref{gp kernel}. The parameters of the two simple harmonic oscillator terms are:

\begin{align}
    Q_1 &= 1/2 + Q_0 + dQ, \\
    Q_2 &= 1/2 + Q_0, \\
    \omega_1 &= \frac{4 \pi Q_1}{P_{\rm rot} \sqrt{4 Q_1^2 - 1}}, \\
    \omega_2 &= \frac{8 \pi Q_1}{P_{\rm rot} \sqrt{4 Q_1^2 - 1}}, \\
    S_1 &= \frac{\sigma_{rot}^2}{(1+f)\omega_1 Q_1}, \\
    S_2 &= \frac{f \sigma_{rot}^2}{(1+f)\omega_2 Q_2}.
\end{align}
\noindent This kernel has two modes in Fourier space: one at period $P$ and one at half the period. We parameterize the kernel by the primary period of variability \Prot{;} the standard deviation of the process $\sigma_{rot}$; the quality factor (minus one half) for the secondary oscillation $Q_0$, which keeps the system underdamped; the difference between the quality factors of the first and second modes $dQ$, which ensures that the primary mode always has higher quality; the fractional amplitude $f$ of the secondary mode compared to the primary; and finally an additive scalar jitter term $s$ is added to account for any excess noise in the activity model. The GP only models correlated noise, so the jitter term is added to capture the uncorrelated noise and is added in quadrature to the RV uncertainty.

The RV Keplerian model is parameterized by the orbital period $P$, the time of mid-transit $T_0$, the log of the semi-amplitude $\ln{K}$, an RV offset $v_0$, and the following reparameterizations of eccentricity $e$ and argument of periastron $\omega$: $h = \sqrt{e}$cos$\omega$, $k = \sqrt{e}$sin$\omega$. Hence, our full RV model consists of our GP and Keplerian models with 12 free parameters: \{\Prot{}, $\sigma_{rot}$, ln$Q_0$, ln$dQ$, $f$, ln$s$, $v_0$, $P$, $T_0$, ln$K$, $h$, $k$\}. The adopted model parameter priors are also included in Table \ref{tab:priors}. We fit the RV data with our full model using the {\tt\string exoplanet} package, which is an extension of the {\tt\string PyMC3} inference engine. Point estimates of the model parameters are derived from their respective marginalized PDFs and are reported in Table \ref{tab:point_estimates}. The point estimates reported represent each parameter's median maximum a-posteriori value and uncertainties from its 16$^{\rm th}$ and 84$^{\rm th}$ percentiles.

Figure \ref{fig:RVs} shows the raw RVs, individual model components, and GLS periodograms. A noticeable but not significant periodicity at \Prot{} $= 48$ days emerges in the GLS periodogram of the RV activity signal, and also at \Prot{}/2 = 24 days to a lesser extent. The GLS periodogram corresponding to the orbital model solution after removing the mean GP activity model is clearly dominated by the 3.13-day periodicity with no other significant periodic signal. We measure an RV semiamplitude of $K = 4.39\pm 0.69$ m s$^{-1}$, which is clearly visible in the phase-folded RVs also shown in Figure \ref{fig:RVs}. The RV residuals show the data minus the mean GP activity model and maximum a-posteriori Keplerian solution, and have an rms of 3.47 m s$^{-1}$ and a reduced $\chi^2$ of 1.39. The GLS periodogram of the residuals lack signals with significant power, which suggests that we do not have evidence for an additional planetary companion in the TOI-1695 system.

\subsection{Search for Additional Transiting Planets}
We also performed a search for additional transiting planets in the TESS light curve using the Transit Least Squares algorithm \citep[\texttt{TLS};][]{Hippke_2019}. We conducted the search on the detrended TESS light curve following the removal of our maximum a-posteriori transit model for TOI-1695 b. We ran separate \texttt{TLS} transit searches on each set of consecutive TESS sectors and over the range of orbital periods from 0.5-30 days. Our \texttt{TLS} search revealed no significant signals indicating that we do not have evidence for repeating transits from an additional planet in the system. We also conducted a close visual inspection of the TESS light curve, which revealed no obvious signature of any single-transit events. Our null detection of a second planet is consistent with the results of the SPOC transit search as indicated in the data validation (DV) report.

\subsection{Transit Timing Variation Search}
We also attempted to measure the individual transit times of  TOI-1695 b in each of the five TESS sectors to search for transit timing variations (TTVs). However, given the low S/N of any  individual transit event in the TESS light curve\footnote{Typical single-transit S/N $=0.55$ for a transit depth of 1.27 ppt compared to a typical photometric rms of 2.31 ppt across the five TESS sectors.}, we were unable to confidently recover 70\% of the individual mid-transit times. For the remaining transit events for which our transit model did converge, we measured the individual $T_{0,i}$ values with a typical uncertainty of $\sim 15$ minutes. These measurements did not reveal any significant TTVs when compared to a linear ephemeris.

\section{Discussion} \label{Discussion}
\subsection{Mass-radius diagram} \label{sect:mass_radius}

\begin{figure}
    \centering
    \includegraphics[width=\hsize]{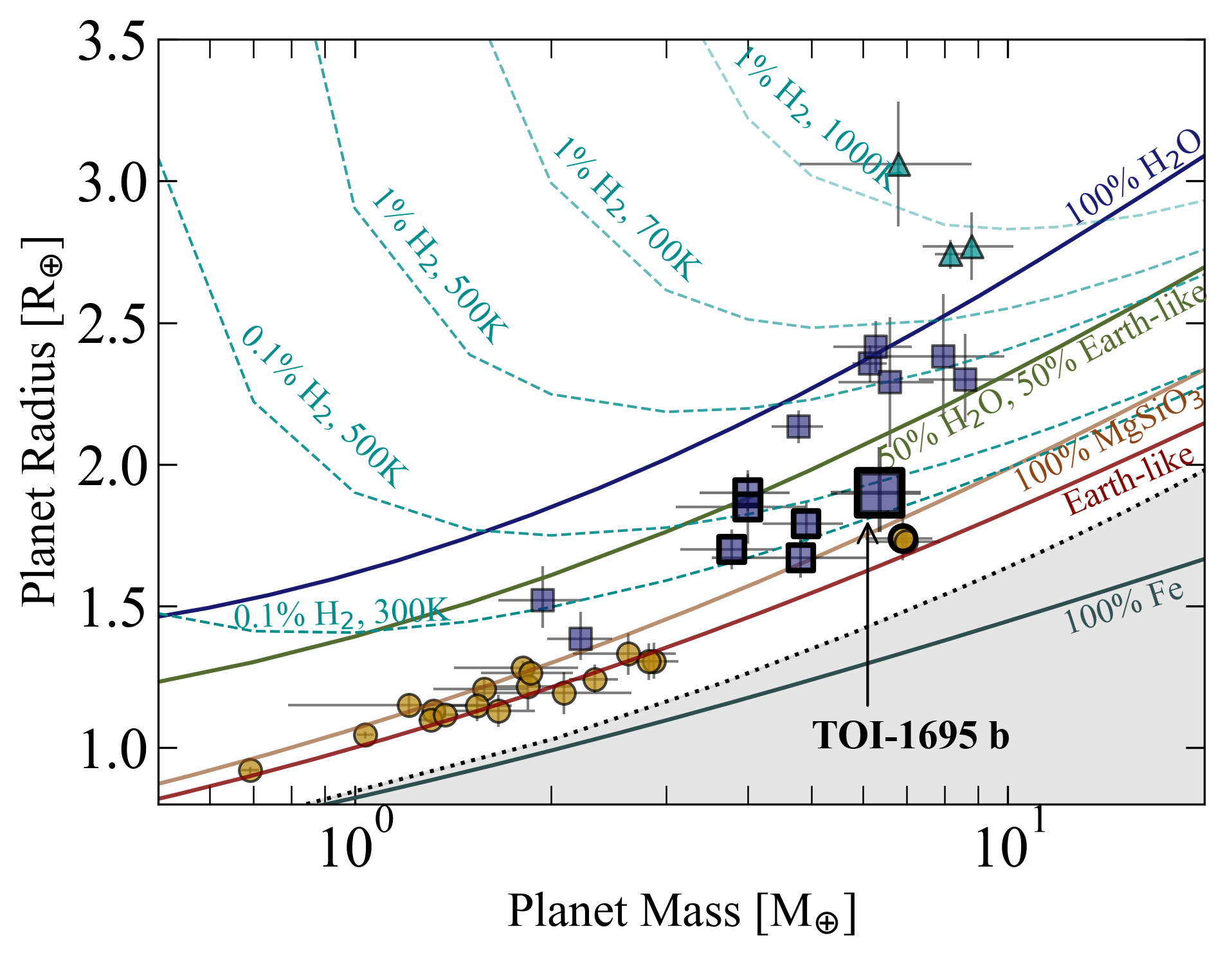}
    \caption{Mass-radius diagram for small planets orbiting M dwarfs with masses measured to better than 3$\sigma$ from NASA Exoplanet Archive, including TOI-1695 b for comparison (large square). Planet marker shapes indicate bulk composition interpretation as rocky (circles), gaseous (triangles), or intermediate (squares). The solid curves depict internal structure models with mass fractions of 100\% water, 33\% iron plus 67\% rock (i.e. Earth-like), and 100\% iron \citep{Zeng_2013}. The dashed curves depict models of Earth-like cores hosting H$_2$ envelopes of varying mass fractions and equilibrium temperatures \citep{Zeng_2019}. The shaded region corresponds to the forbidden region according to models of maximum collisional mantle stripping by giant impacts \citep{Marcus_2010}. The seven keystone planets are bolded.}
    \label{fig:mass_radius}
\end{figure}

Our analysis of the {\tt\string PDCSAP} light curve reveals that TOI-1695 b has an orbital period of $P = 3.1342791^{+0.0000071}_{-0.0000063}$ days. Using the stellar parameters presented in Table~\ref{tab:stellar_params}, we find that the semimajor axis of the planetary orbit is $a = 0.033516^{+0.000087}_{-0.000086}$ au where it receives an instellation flux of $F = 39\pm 3$ $F_\oplus$. Assuming uniform heat redistribution and a Bond albedo of zero, the corresponding equilibrium temperature of TOI-1695 b is $T_{eq} = 698\pm 14$ K.

We also measure a planetary radius and mass of $R_p = 1.90^{+0.16}_{-0.14}\ R_\oplus$ and $M_p = 6.36 \pm 1.00\ M_\oplus$, corresponding to 12.7$\sigma$ and 6.4$\sigma$ detections, respectively. The mass and radius measurements yield a 2.8$\sigma$ bulk density measurement of $\rho_p = 5.0^{+1.8}_{-1.3}$ g cm$^{-3}$. 

Figure \ref{fig:mass_radius} compares the mass and radius of TOI-1695 b to the population of small M dwarf planets with masses measured to better than 3$\sigma$, which we retrieve from the NASA Exoplanet Archive. Planets are classified based on their bulk compositions inferred from their mass and radius measurements. Earth-like planets are defined as those consistent with an Earth-like composition curve, gas-rich planets cannot be explained by even 100\% water composition and require an extended H/He envelope, and the remaining planets we broadly classify as ``intermediate'' given that their masses and radii are between those of the aforementioned groups. We find that the bulk composition of TOI-1695 b is inconsistent with that of the Earth at 1.7$\sigma$. Specifically, TOI-1695 b is underdense relative to an Earth-like composition of the same mass. The planet could therefore belong to the population of enveloped terrestrials whose rocky components resemble Earth but require an extended gaseous envelope to explain their masses and radii. Such is the expected composition of the majority of sub-Neptunes predicted by thermally-driven mass loss models \citep{Owen_2017, Gupta_2019}. In Section \ref{photoevaporation}, we show that TOI-1695 b would require an envelope mass fraction of 0.06\% to explain its mass and radius. Assuming a H/He envelope with solar metallicity (mean molecular mass = 2.35 $u$), we also show that the planet is unlikely to retain such an envelope over long timescales since it is susceptible to thermally-driven hydrodynamic escape at 39 times Earth insolation. Thus, it is unlikely to be an enveloped terrestrial with a primordial H/He atmosphere. Instead, it is likely that TOI-1695 b possesses a high mean molecular weight envelope supplied by volatile delivery, or formation beyond the ice line followed by migration and volatile retention. We find that the mass and radius of TOI-1695 b is consistent with a water mass fraction of $31^{+33}_{-22}$\% assuming a two layer model of MgSiO$_3$ and H$_2$O. This composition is consistent within 1$\sigma$ of the subpopulation of water worlds with water mass fraction of 50\% \citep[See Section \ref{TDML};][]{Luque_2022}.

\subsection{Photoevaporation model: limits on envelope mass fraction}
\label{photoevaporation}

\begin{figure}
    \centering
    \includegraphics[width=\hsize]{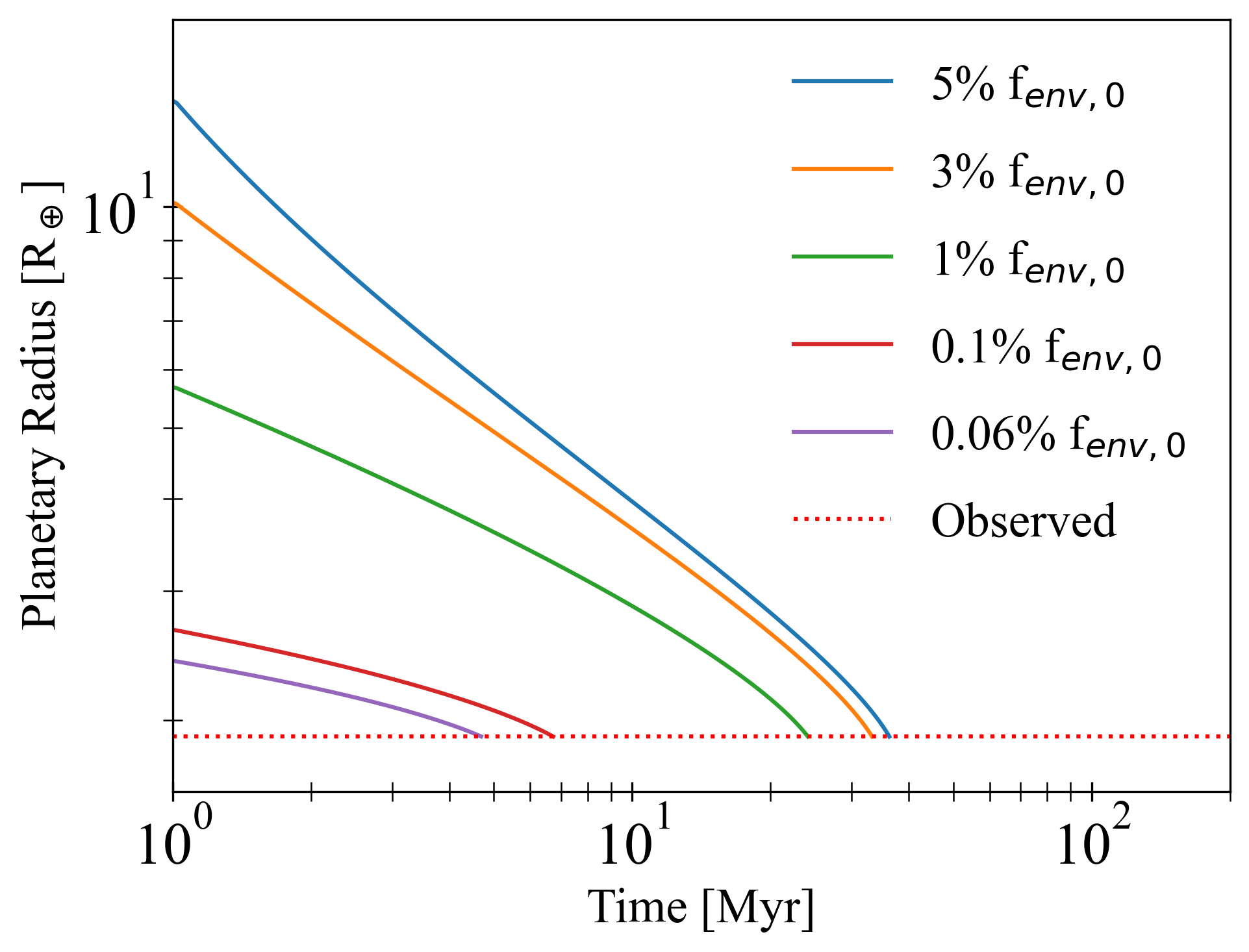}
    \caption{Results of our photoevaporation simulation, which models EUV-driven hydrodynamic escape and thermal contraction on TOI-1695 b. Each curve corresponds to a different initial envelope mass fraction. The planetary radius decreases to the observed radius (1.90 $R_\oplus$) on Myr timescales, making present day H/He enveloped rocky core an unlikely composition.}
    \label{fig:photoevaporation}
\end{figure}

\begin{figure}
    \centering
    \includegraphics[width=1.1\hsize]{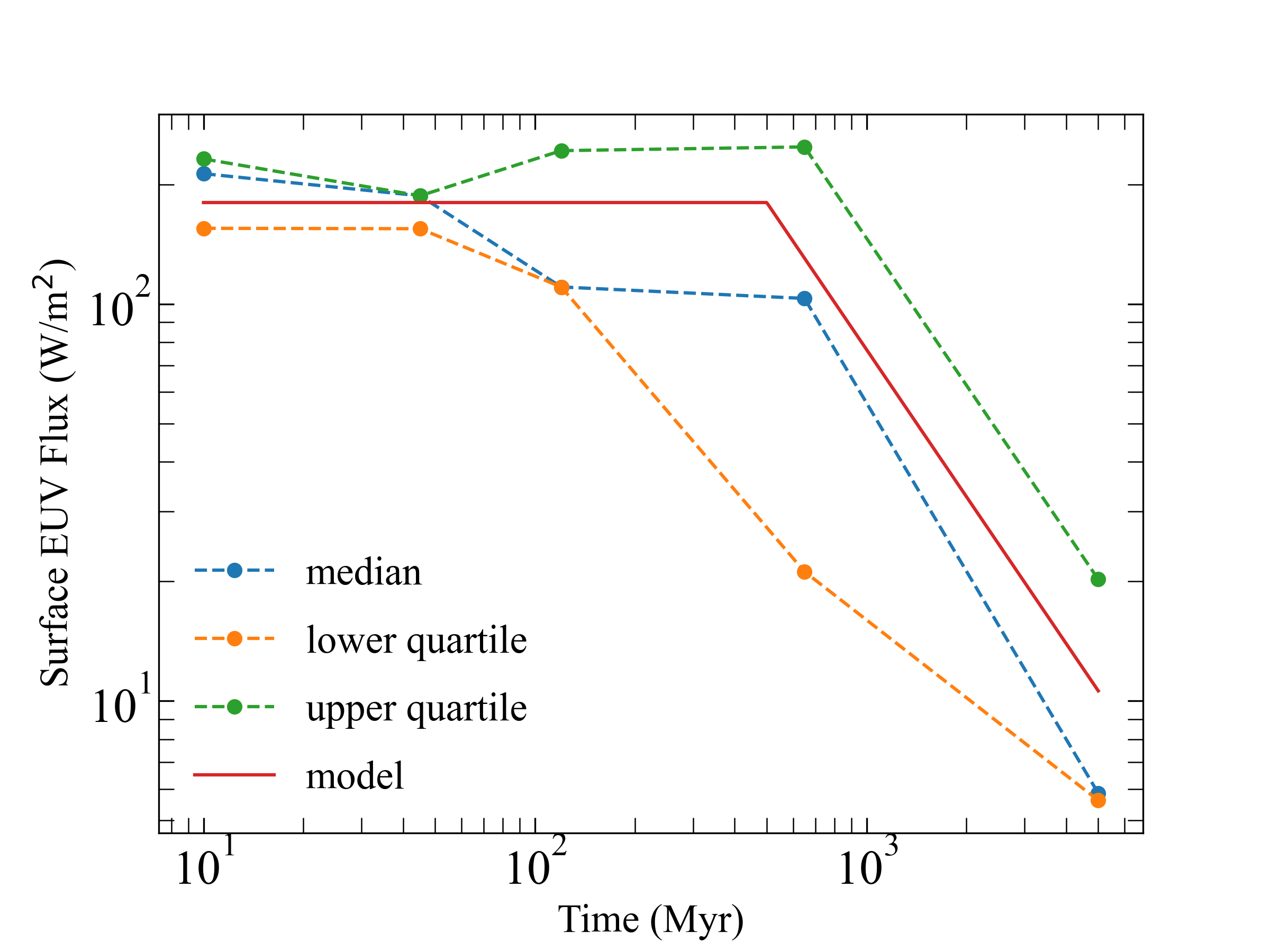}
    \caption{The solid red line shows the stellar surface EUV flux evolution model used in the TOI-1695 b photoevaporation simulations (Section \ref{photoevaporation}, Equation \ref{eq:flux}). The dashed lines show the integrated flux density for 10 nm $<$ $\lambda$ $<$ 130 nm from the HAZMAT synthetic spectra \citep{Peacock_2020}. These spectra are computed for the lower quartile, median, and upper quartile EUV flux density samples of early M dwarfs at five ages: 10 Myr, 45 Myr, 120 Myr, 650 Myr, and 5 Gyr. We chose $F_{EUV, 0}$ = 180 W m$^{-2}$ for our model. Our EUV flux model is shown to be consistent with the semi-empirical data.}
    \label{fig:EUV_model}
\end{figure}

We ran photoevaporation simulations to assess the possibility that TOI-1695 b could be an Earth-like core enveloped by a H/He-dominated atmosphere. Our model assumes extreme ultraviolet (EUV; 10 nm $< \lambda <$ 130 nm)-driven hydrodynamic escape is the primary driver of mass loss. We opt to ignore X-ray irradiation since approximately 80-95\% of the high energy flux is in the EUV and to remain consistent with available synthetic spectra discussed below \citep{Loyd_2016, Fontenla_2016, Peacock_2019, Peacock_2019b, Peacock_2020}. We also assume a H/He composition with solar metallicity (mean molecular mass = 2.35 $u$) and assume that hydrogen is atomic due to photodissociation.

In order to model the time evolution of the incident EUV flux, we constructed a power law function of the EUV flux's temporal evolution. Our model is informed by semi-empirical spectra generated by the HAZMAT team for populations of M1 stars over a range of ages from 10 Myr to 5 Gyr \citep[Figure \ref{fig:EUV_model};][]{Peacock_2020}:

\begin{equation}
    F_{EUV} = 
    \begin{cases}
    F_{EUV, 0} & t < t_{sat} \\
    F_{EUV, 0} \left(\frac{t}{t_{sat}} \right)^\beta & t > t_{sat},
    \end{cases}
\label{eq:flux}
\end{equation}

\noindent where $t_{sat}$ is the saturation time marking the transition from constant EUV flux phase to power law decay and $F_{EUV}$ is incident planetary EUV flux (related to stellar EUV luminosity $L_{EUV}$ by: $F_{EUV} = L_{EUV}/4 \pi a^2$, where $a$ is the orbital semi-major axis). We found $\beta = -1.23$ and $t_{sat} = 500$ Myr to be consistent with the HAZMAT spectra as well as previously reported solar data \citep{Ribas_2005}.

EUV-driven escape generates a mass flux, which we compute as a function of an efficiency factor $\epsilon$, incident EUV flux $F_{EUV}$, and the planetary gravitational potential $V_{pot}$:

\begin{equation}
    \phi = \frac{\epsilon F_{EUV}}{4 V_{pot}},
    \label{eq:massflux}
\end{equation}
where $V_{pot} = GM_p/R_p^2$. $\epsilon$ encapsulates several heat transfer processes, ultimately representing the fraction of incident radiation that drives escape. We chose a value of $\epsilon$ = 0.15, consistent with previously reported lower estimates \citep{Watson_1981, Schaefer_2016}.

The evolution of the planetary radius through time was determined by a combination of atmospheric escape described by $\phi$ and contraction due to cooling described by thermal evolution models \citep{Lopez_2014}. Our simulations demonstrate that TOI-1695 b loses its entire atmosphere on rapid timescales of the order 1-100 Myr for a wide range of initial atmospheric mass fraction values (0.06\% - 5\%; Figure \ref{fig:photoevaporation}). This finding is robust over a wide range of values for $t_{sat}$ and $\epsilon$. We find that the observed planetary radius is consistent with a 0.06\% H/He-dominated atmosphere, which the planet could not have retained over reasonably observable periods of time. We conclude that a present-day H/He-dominated atmosphere is highly unlikely, barring a steady state outgassing scenario.

\subsection{Keystone planets and implications for the M dwarf radius valley} \label{sect:keystone}

\begin{figure}
    \centering
    \includegraphics[width=\hsize]{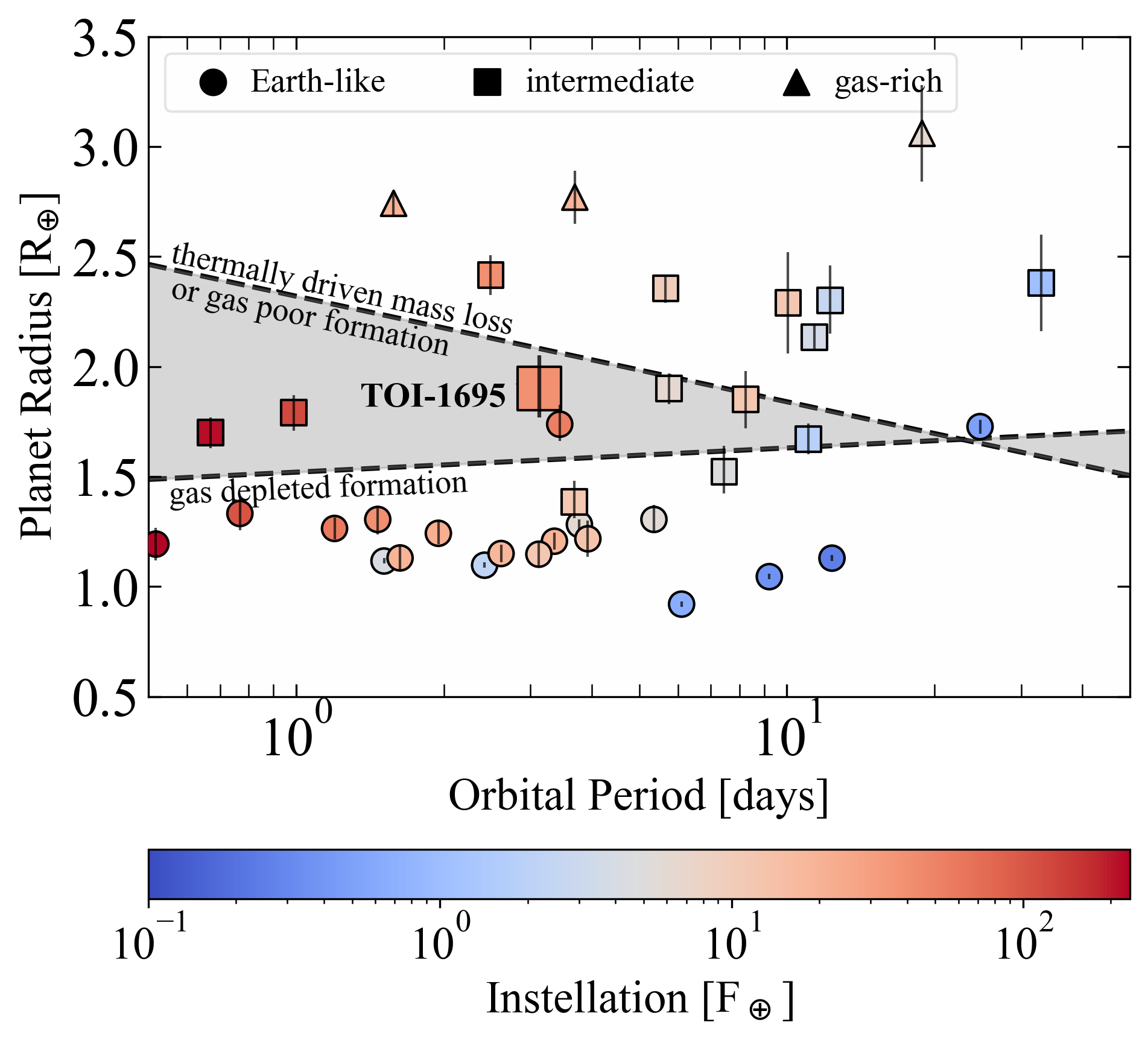}
    \caption{Period–radius diagram for small planets transiting M dwarfs and with precisely measured RV masses ($>3\sigma$). The dashed lines depict model predictions of the location of the M dwarf radius valley from thermally-driven mass loss and from gas depleted formation. The shaded wedge regions host the so-called keystone planets, including the newly discovered TOI-1695 b. The marker shapes depict planets whose bulk compositions have been determined to be Earth-like (circles), gas-rich (triangles), or intermediate (squares; see Section~\ref{sect:mass_radius} for definitions). The color bar highlights each planet’s incident instellation.}
    \label{fig:period_radius}
\end{figure}


Several physical mechanisms have been proposed to explain the emergence of the radius valley around FGK and late-K to mid-M dwarfs, each making distinct predictions of the slope of the radius valley in period-radius space parameterized as $R_{p,\text{valley}} \propto P^\beta$. These include two thermally-driven atmospheric mass loss models: stellar XUV-driven hydrodynamic escape in which atmospheric species flow outward in the form of a Parker wind \citep{Owen_2013, Jin_2014, Lopez_2014, Chen_2016, Jin_2018}; and core-powered mass loss in which the planetary core's formation energy drives escape over Gyr timescales \citep{Ginzburg_2018, Gupta_2019, Gupta_2020}. An alternative mechanism is the gas poor (but not gas depleted) formation scenario which suggests a primordial radius valley in which gas accretion onto low mass cores is limited \citep[$\lesssim 1-2\ M_\oplus$;][]{Lee_2021,Lee_2022}. These three models predict negative values of $\beta \in [-0.15, -0.09]$ \citep{Lopez_2018, Gupta_2020, Lee_2021}. Yet a fourth model supposes that enveloped terrestrials form within the first several Myr when the gaseous disk is still present, whereas terrestrial planets form at later times after the dissipation of the gaseous disk in a gas depleted environment. The gas depleted model predicts the opposite sign for the period dependence of the radius valley \citep[$\beta = 0.11$;][]{Lopez_2018}.

The slope of the radius valley around Sun-like stars with \teff\ $>$ 4700 K has been well characterized and measurements of $\beta$ take values of $[-0.11, -0.09]$ \citep{Eylen_2018, Martinez_2019}, consistent with thermally-driven mass loss and gas poor formation model predictions. However, in the lower stellar mass regime of late-K to mid-M dwarfs, tentative evidence suggests that $\beta = 0.06 \pm 0.02$, which is inconsistent with the values measured around FGK stars and is instead consistent with predictions from gas depleted formation models \citep{Cloutier_2020a}. 


While occurrence rate measurements around low mass stars are currently insufficient to resolve a transition in competing radius valley emergence mechanisms between the low stellar mass and high stellar mass regime, we may gain insight by characterizing keystone planets like TOI-1695 b, which span the model predictions in period-radius space. The distinct slopes of the radius valley's period dependence carve out a wedge in the orbital period-planet radius space, between which the competing models make conflicting predictions (shaded region in Figure \ref{fig:period_radius}). At periods less than 23.5 days, thermally-driven mass loss and gas poor formation models predict that planets in the wedge (i.e. keystone planets) are rocky. Conversely, gas depleted formation models predict they should host gaseous envelopes because their size exceeds the maximum rocky planet mass that can form out of the minimum mass extrasolar nebula at its observed orbital separation. At $P$ = 3.13 days and $R_p$ = 1.90$^{+0.16}_{-0.14}$ $R_\oplus$, TOI-1695 b is one such keystone planet, whose composition directly constrains the prominence of the competing physical processes on close-in planets around early M dwarfs (Figure \ref{fig:period_radius}). 

Figure \ref{fig:period_radius} also features the same population of planets displayed in Figure \ref{fig:mass_radius}. Intermediate planets may be explained by a variety of compositions including a H/He envelope, a volatile-rich composition, or perhaps an exotic rocky composition that is enhanced in Ca and Al \citep{Dorn_2018}. Our analysis reveals that TOI-1695 b is inconsistent with Earth-like and gas-rich compositions and requires an alternative physical interpretation to explain its mass and radius. The bulk composition of TOI-1695 b is therefore inconsistent with predictions from thermally-driven mass loss models. 


So what about the prospect that TOI-1695 b formed via a gas depleted formation scenario? If the gas depleted formation mechanism were operating in the TOI-1695 system, then TOI-1695 b must have formed early on before disk dispersal and subsequently accreted - and at least partially retained - a primordial H/He envelope. However, if TOI-1695 b accreted its primordial envelope at its current location, then such an atmosphere should have been rapidly lost to thermal escape, as demonstrated in Section \ref{photoevaporation}. This discrepancy may be reconciled if the planet hosts a high mean molecular weight atmosphere and/or migrated inward to its current location so as to avoid the bulk stellar XUV output during the first few hundreds of Myrs. We therefore conclude that the classical picture of the gas depleted formation model, which only produces gas-enveloped terrestrials and terrestrial cores that are born rocky, cannot explain the observed composition of TOI-1695. Instead, we speculate that TOI-1695 b is more likely to be rich in volatiles. However, we emphasize that our conclusion that TOI-1695 b is inconsistent with a thermally-driven mass loss scenario remains.

\subsection{How likely is the keystone planet population being sculpted by a thermally-driven mass loss process?} 
\label{TDML}

At the time of this publication, there are seven M dwarf keystone planets for which reliable mass and radius measurements are available (Table~\ref{tab:keystone}). Using this sample, we ask the question: what is the probability that the keystone planet population around M dwarfs is sculpted by a thermally-driven mass loss process ($P(\mathrm{TDML})$)? To answer this question, we must first define the probability that the composition of each keystone planet is consistent with a thermally-driven mass loss hypothesis ($P_i$). For this we compute the probability that each planet is consistent with having an Earth-like composition based on its mass and radius. We compute each $P_i$ as the fraction of samples from the planet's joint mass-radius posterior that result in a sampled radius that is less than the radius of a pure MgSiO$_3$ planet at the sampled mass value. This criterion adopted to define the radius upper limit for an Earth-like planet is equivalent to our definition described in Section \ref{sect:mass_radius} and shown in Figure~\ref{fig:mass_radius}. Our probabilities $P_i$ are included in Table~\ref{tab:keystone}.

Although the probability that any individual keystone planet is consistent with thermally-driven mass loss is often not very illuminating, the statistical statement that we can make from the seven keystone planet sample is meaningful. By treating the measurement of each keystone planet's mass and radius as an independent Bernoulli experiment, we can calculate the probability that the keystone planet population is being sculpted by thermally-driven mass loss as the product of the individual probabilities (i.e. $P(\mathrm{TDML}) = \prod_{i=1}^7 P_i$). That is, what is the probability that all seven keystone planets have Earth-like compositions and are therefore consistent with a thermally-driven mass loss scenario. We find that $P(\mathrm{TDML}) = 5.9\times 10^{-10}$. Similarly, we evaluate the probability that the keystone planet population is {\it inconsistent} with a thermally-driven mass loss scenario and find that $P(\mathrm{not\ TDML}) = \prod_{i=1}^7 (1 - P_i) = 9.4\times 10^{-2}$. Comparing these values, we find $P(\mathrm{not\ TDML})/P(\mathrm{TDML}) = 1.6\times 10^{8}$. Thus, the M dwarf keystone planet population strongly disfavors a thermally-driven mass loss scenario. This result is not surprising as only one planet out of seven (i.e. TOI-1235 b) has a greater than 50\% chance of having an Earth-like composition and therefore with being consistent with a thermally-driven mass loss process.

Our results are consistent with the emerging picture that the M dwarf radius valley is a by-product of planet formation and is not sculpted by thermally-driven mass loss. Early investigations of close-in planet occurrence rates around late-K to mid-M dwarfs suggested that the slope of the radius valley with instellation was inconsistent with predictions from thermally-driven mass loss models \citep{Cloutier_2020a}. More recent empirical evidence from \cite{Luque_2022} demonstrated that the sub-Neptune peak in the M dwarf radius valley represents water-rich planets and not gas-enveloped terrestrials. \cite{Luque_2022} refined the masses and radii of small transiting planets around M dwarfs and revealed three distinct planet types: Earth-like, water worlds, and puffy sub-Neptunes. The mass-radius profiles of these subpopulations are consistent with Earth-like, 50\% water-dominated ices/50\% silicates, and H/He-enveloped compositions respectively, and are interpreted as such. It is concluded that rocky planets must form within the water ice line while water worlds form beyond and migrate inward. In this case, the apparent radius valley around M dwarfs is sculpted by accretion history rather than by atmospheric mass loss. If water worlds are indeed ubiquitous, it is likely that TOI-1695 b belongs to this subpopulation. Our calculated water mass fraction of $31^{+33}_{-22}$ \% supports this interpretation and is consistent with the 50\% water mass fraction subpopulation from \cite{Luque_2022}. Taken together, the slope of the M dwarf radius valley \citep{Cloutier_2020a}, the recovery of a population of likely water-rich planets \citep{Luque_2022}, and our results for keystone planets suggest that thermally-driven mass loss does not explain the origin of the M dwarf radius valley. Instead, it is likely to emerge directly from the planet formation process.


\input{keystone_planet_table}

\subsection{RV sensitivity: limits on additional planets}

\begin{figure}
    \centering
    \includegraphics[width=\hsize]{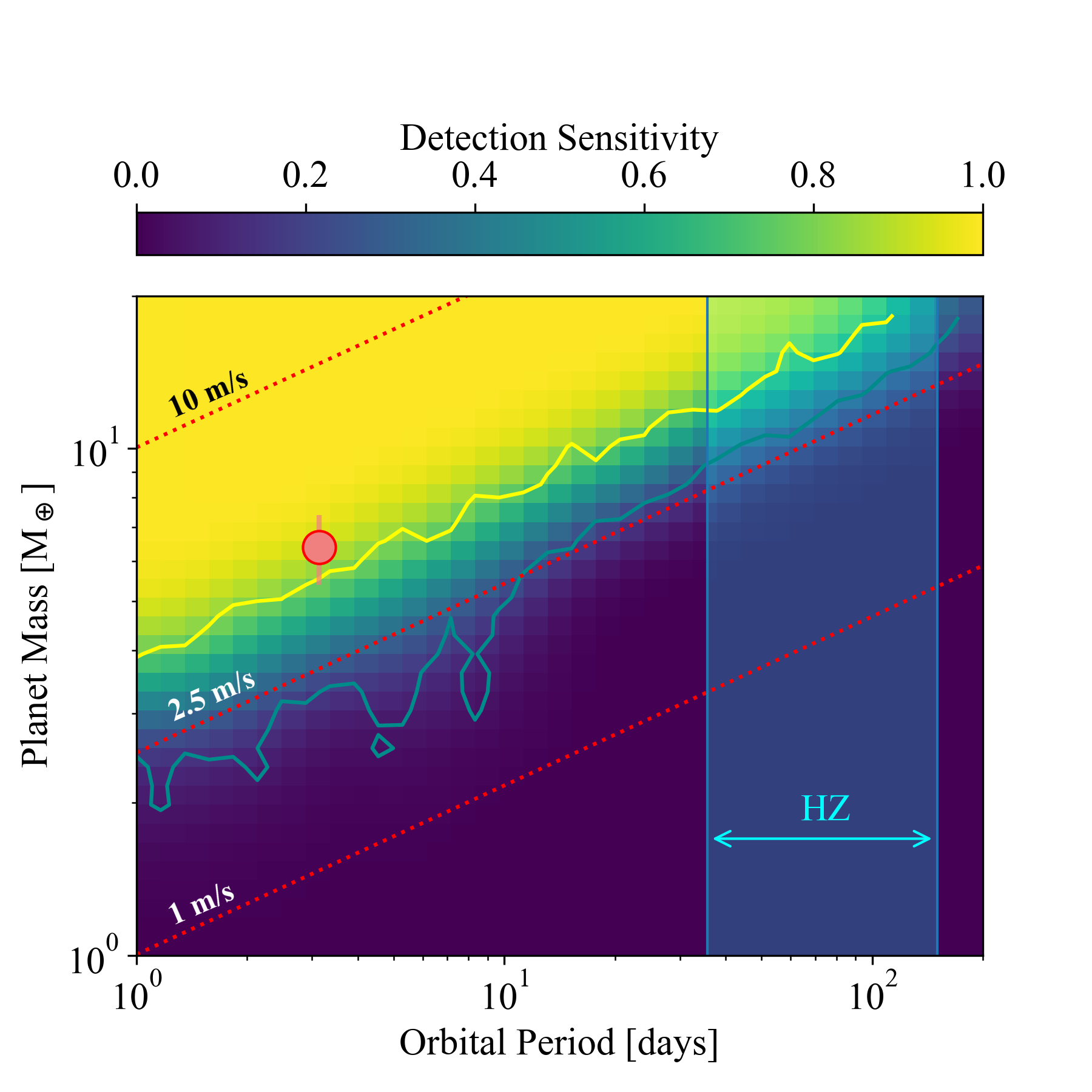}
    \caption{RV detection sensitivity to planets orbiting TOI-1695 as a function of planet mass and orbital period. The solid lines mark the 10\% and 90\% sensitivity limits, respectively. The circle marker highlights TOI-1695 b. Dotted red lines show constant semi-amplitudes of 1, 2.5, and 10 m/s. The shaded region spans the habitable zone of TOI-1695 whose inner and outer edges are defined by the recent Venus and early Mars boundaries \citep{Kopparapu_2013}.}
    \label{fig:RV_sensitivity_plot}
\end{figure}

Exoplanet transit surveys and RV follow-up of transiting systems have shown that multi-planet systems are common around M dwarfs \citep{Dressing_2015, Gaidos_2016, Cloutier_2021b}. It is therefore reasonable to expect additional planets in the TOI-1695 system that evade detection due to their small sizes, long orbital periods, or non-transiting orbital configurations. We assessed the detection sensitivity of our HARPS-N RV dataset to place constraints on the presence of additional planets by computing our detection sensitivity as a function of orbital period and planet mass via a set of injection-recovery tests. We took a Monte Carlo approach by injecting synthetic Keplerian signals into the residuals of the HARPS-N RV time series after removing of the maximum a-posteriori RV solution (i.e. TOI-1695 b plus GP). We inject a single planet in each of the 10$^5$ iterations. To generate the Keplerian signal for each iteration, planet masses and orbital periods were sampled uniformly in log space from 1 - 20 $M_\oplus$ and 1 - 200 days, respectively. Note that our RV baseline is 428 days. Because compact multi-planet systems are often nearly co-planar, we sampled orbital inclinations from a Gaussian distribution $\mathcal{N}$($i_b$, $\sigma_i$), where $i_b$ = 88.5$^\circ$ and we adopt the dispersion of mutual inclinations of $\sigma_i$ = 2$^\circ$ following from studies of multi-planet M dwarf systems \citep{Ballard_2016}. We sampled the stellar mass from its posterior and used it to calculate the RV semiamplitude assuming a circular orbit. We injected the resulting Keplerian signals into the RV residuals while preserving the individual measurement uncertainties and timestamps.

Recovery of the injected synthetic planets corresponds to a successful detection and involved a two step process. To warrant a detection, an injected signal must first produce a significant peak in a GLS periodogram with a false alarm probability (FAP) of $\leq 1\%$. The GLS periodogram was constructed for each iteration and the analytical FAP was calculated using the analytical formalism described by \cite{Zechmeister_2009}. Second, the six-parameter Keplerian model must be strongly favored over the null hypothesis (i.e. a flat line with a constant offset). To perform the model comparison, we calculated the Bayesian information criterion (BIC) for each model as BIC $= 2\ln{\mathcal{L}} + x \ln{N}$, where $\mathcal{L}$ is the likelihood of the RV data given the assumed model, $x$ is the number of model parameters (i.e. 1 and 6 for the null and Keplerian models, respectively), and $N = 49$ is the number of RV measurements. We claimed successful recovery of an injected planet signal if both criteria are met: the GLS periodogram power of the largest signal within $\pm$ 2\% of the injected period has FAP $\leq$ 1\%, and $\Delta$BIC = BIC$_{Keplerian}$ - BIC$_{null}$ $\geq$ 10. The sensitivity of our RV dataset is defined as the ratio of recovered planets to injected planets and is presented in Figure \ref{fig:RV_sensitivity_plot}.

Figure \ref{fig:RV_sensitivity_plot} shows that the mass and orbital period of TOI-1695 b lies above the 90\% detection contour. We also find that we are sensitive to approximately 50\% of planets at 3 $M_\oplus$ and 90\% of planets at 4 $M_\oplus$ at a 1-day orbital period. Within 10 days, we are sensitive to all planets $\geq 10 M_\oplus$. Adopting the empirical recent Venus and early Mars habitable zone (HZ) limits from \cite{Kopparapu_2013}, i.e. 35 - 150 days, we find that we are only sensitive to very massive HZ planets ($> 15\ M_\oplus$). Such planets would likely host massive gaseous envelopes, rendering their surfaces uninhabitable by the traditional definition of the HZ.

Additionally, we performed a blind search over a wide period space with {\tt\string RVSearch}, which revealed no significant signals below the 0.1\% FAP threshold \citep{Rosenthal_2021}. We conclude that additional planetary signals beyond TOI-1695 b are not detectable in our RV data.

\subsection{An Independent Analysis of the TOI-1695 System}

Following the announcement of the TOI-1695.01 level-one planet candidate, multiple precise RV instrument teams began following up this target through the TESS Follow-up Observing Program (TFOP). In this study we have presented the efforts from the HARPS-N Collaboration and we acknowledge that another collaboration is also in the process of presenting their own RV time series and analysis using data from the SPIROU near-IR spectropolarimeter (Kiefer et al. in prep). While the submissions of these complementary studies were coordinated between the two groups, their respective data, analyses, and write-ups were intentionally conducted independently.

\section{Summary} \label{Summary}

We presented the discovery of TOI-1695 b, a keystone planet orbiting an M1 dwarf. We characterized the planet using TESS transit data and HARPS-N follow up RVs. Keystone planet bulk composition characterization is useful for distinguishing between prevailing radius valley emergence models of thermally-driven mass loss versus gas depleted formation. Our main findings include:

\begin{enumerate}
    \item TOI-1695 b is a sub-Neptune planet with $P = 3.1342791^{+0.0000071}_{-0.0000063}$ days, $R_p = 1.90^{+0.16}_{-0.14}\ R_\oplus$, and $M_p = 6.36 \pm 1.00\ M_\oplus$. The exact bulk composition of TOI-1695 b is degenerate but is notably underdense relative to an Earth-like composition.
    
    \item Our photoevaporation model demonstrates that it is highly unlikely for TOI-1695 b to have retained a H/He envelope. We conclude that the most likely composition is an Earth-like rocky component with a substantial water-rich volatile component. The planetary mass and radius are consistent with a MgSiO$_3$/H$_2$O bilayer with a water mass fraction of $31^{+33}_{-22}$ \%, consistent with the water world subpopulation reported by \cite{Luque_2022} to 1$\sigma$.
    
    \item The bulk composition of TOI-1695 b is inconsistent with predictions from photoevaporation, core-powered mass loss and gas poor formation mechanisms. As such, TOI-1695 b supports the emerging idea that the population of planets within the radius valley around M dwarfs with masses $\lesssim 0.6\ M_\odot$ may not be sculpted by a thermally-driven mass loss process.
    
    \item TOI-1695 b becomes the seventh well-characterized keystone planet around an early M dwarf. As only one out of seven keystone planets (TOI-1235 b) are likely consistent with photoevaporation, core-powered mass loss, and gas poor formation mechanisms, we showed that this planet sample strongly disfavors a thermally-driven mass loss scenario by a factor of $1.2\times 10^{8}$.
    
    \item Along with evidence from \cite{Cloutier_2020a} showing that the M dwarf radius valley slope with period is inconsistent with thermally-driven mass loss, and evidence from \cite{Luque_2022} that the sub-Neptune peak represents water-rich planets, our finding that the keystone planet population is inconsistent with thermally-driven mass loss marks the third major piece of evidence that the M dwarf radius valley emerges as a direct by-product of planetary formation. That is, the M dwarf radius valley likely reflects a distribution of planets that are born rocky, volatile-rich, or gas-enveloped, rather than being sculpted by thermally-driven mass loss.
    
\end{enumerate}

\facilities{ESA/Gaia, NASA/TESS, ASAS-SN, FLWO/TRES, MLO, LCOGT, Keck/NIRC2, TNG/HARPS-N}

\software{{\tt AstroImageJ} \citep{Collins:2017},
        {\tt astropy} \citep{Astropy_2013, Price_2018},
        {\tt astroquery} \citep{astroquery},
        {\tt BANZAI} \citep{McCully:2018},
        {\tt celerite} \citep{ForemanMackey_2017},
        {\tt exoplanet} \citep{ForemanMackey_2019},
        {\tt Lightkurve} \citep{lightkurve},
        {\tt PyMC3} \citep{Salvatier_2016},
        {\tt RVSearch} \citep{Rosenthal_2021},
          {\tt STARRY} \citep{Luger_2019},
          {\tt TERRA} \citep{Anglada_2012}.
          }

\vspace{1cm}
We thank Evgenya Shkolnik for fruitful discussions of the high energy radiation of M dwarfs and Robin Wordsworth for his input on the photoevaporation model.

RC is supported by the Banting Postdoctoral Fellowship Program administered by the Government of Canada. 

This paper includes data collected by the TESS mission. Funding for the TESS mission is provided by the NASA's Science Mission Directorate.

We acknowledge the use of public TESS data from pipelines at the TESS Science Office and at the TESS Science Processing Operations Center.

Resources supporting this work were provided by the NASA High-End Computing (HEC) Program through the NASA Advanced Supercomputing (NAS) Division at Ames Research Center for the production of the SPOC data products.

Partly based on observations made with the Italian {\it Telescopio Nazionale
Galileo} (TNG) operated by the {\it Fundaci\'on Galileo Galilei} (FGG) of the
{\it Istituto Nazionale di Astrofisica} (INAF) at the
{\it  Observatorio del Roque de los Muchachos} (La Palma, Canary Islands, Spain).

This research has made use of the NASA Exoplanet Archive, which is operated by the California Institute of Technology, under contract with the National Aeronautics and Space Administration under the Exoplanet Exploration Program. The data can be accessed via \dataset[NASA Exoplanet Archive]{\doi{10.26133/NEA1}}.

This work makes use of observations from the LCOGT network. Part of the LCOGT telescope time was granted by NOIRLab through the Mid-Scale Innovations Program (MSIP). MSIP is funded by NSF.

This work has made use of data from the European Space Agency (ESA) mission {\it Gaia} (\url{https://www.cosmos.esa.int/gaia}), processed by the {\it Gaia} Data Processing and Analysis Consortium (DPAC, \url{https://www.cosmos.esa.int/web/gaia/dpac/consortium}). Funding for the DPAC has been provided by national institutions, in particular the institutions participating in the {\it Gaia} Multilateral Agreement.

We acknowledge the following TESS Guest Investigator (GI) programs that observed TOI-1695 at 2-minute cadence: G022198 – PI Courtney Dressing; G04039 – PI James Davenport; G04148 – PI Paul Robertson; G04178 – PI Joshua Pepper; G04242 – PI Andrew Mayo.

This material is based upon work supported by the National Aeronautics and Space Administration under grants 80NSSC22K0166 and 80NSSC22K0296 in support of the TESS Guest Investigator Program.

B.S.S. acknowledges the support of Ministry of Science and Higher Education of the Russian Federation under the grant 075-15-2020-780 (N13.1902.21.0039).



This research made use of Lightkurve, a Python package for Kepler and TESS data analysis \citep{lightkurve}.

Some of the data presented in this paper were obtained from the Mikulski Archive for Space Telescopes (MAST) at the Space Telescope Science Institute. The specific observations analyzed can be accessed via \dataset[MAST]{\doi{10.17909/t9-st5g-3177}}.


\suppressAffiliationsfalse
\allauthors

\bibliographystyle{aasjournal}
\bibliography{refs.bib}

\end{document}

%% file: authors.tex
\author[0000-0002-8466-5469]{Collin Cherubim}
\affiliation{Earth and Planetary Science, Harvard University, 20 Oxford St, Cambridge, MA 02138, USA}
\affiliation{Center for Astrophysics \textbar \ Harvard \& Smithsonian, 60 Garden Street, Cambridge, MA 02138, USA}

\author[0000-0001-5383-9393]{Ryan Cloutier}
\altaffiliation{Banting Fellow}
\affiliation{Dept. of Physics \& Astronomy, McMaster University, 1280 Main St West, Hamilton, ON L8S 4L8, Canada}
\affiliation{Center for Astrophysics \textbar \ Harvard \& Smithsonian, 60 Garden Street, Cambridge, MA 02138, USA}

\author[0000-0002-9003-484X]{David Charbonneau}  
\affiliation{Center for Astrophysics \textbar \ Harvard \& Smithsonian, 60 Garden Street, Cambridge, MA 02138, USA}

\author[0000-0002-5402-9613]{Bill Wohler}
\affiliation{SETI Institute, Mountain View, CA 94043, USA}
\affiliation{NASA Ames Research Center, Moffett Field, CA 94035, USA}

\author[0000-0003-2163-1437]{Chris Stockdale}
\affiliation{Hazelwood Observatory, Australia}

\author[0000-0002-3481-9052]{Keivan G.\ Stassun}
\affiliation{Department of Physics and Astronomy, Vanderbilt University, Nashville, TN 37235, USA}

\author[0000-0001-8227-1020]{Richard P. Schwarz}
\affiliation{Center for Astrophysics \textbar \ Harvard \& Smithsonian, 60 Garden Street, Cambridge, MA 02138, USA}

\author[0000-0003-1713-3208]{Boris Safonov}
\affiliation{Sternberg Astronomical Institute, Lomonosov Moscow State University, 119992, Moscow, Russia, Universitetskii prospekt, 13}

\author[0000-0001-7254-4363]{Annelies Mortier}  
\affiliation{KICC \& Cavendish Laboratory, University of Cambridge,
  J.J. Thomson Avenue, Cambridge CB3 0HE, UK}
\affiliation{School of Physics \& Astronomy, University of Birmingham, Edgbaston, Birmingham B15 2TT, UK}
  
\author[0000-0003-0828-6368]{Pablo Lewin}
\affiliation{The Maury Lewin Astronomical Observatory, Glendora, CA 91741, USA}

\author{David W. Latham}  
\affiliation{Center for Astrophysics \textbar \ Harvard \& Smithsonian, 60 Garden Street, Cambridge, MA 02138, USA}

\author{Keith Horne}
\affiliation{SUPA Physics and Astronomy, University of St. Andrews, Fife, KY16 9SS Scotland, UK}

\author{Rapha\"{e}lle D. Haywood}  
\affiliation{Astrophysics Group, University of Exeter, Exeter EX4 2QL, UK}
  
\author{Erica Gonzales}  
\affiliation{Department of Astronomy and Astrophysics, University of California, Santa Cruz, CA 95064, USA}

\author{Maria V. Goliguzova} 
\affiliation{Sternberg Astronomical Institute, Lomonosov Moscow State University, 119992, Moscow, Russia, Universitetskii prospekt, 13}

\author[0000-0001-6588-9574]{Karen A.\ Collins}
\affiliation{Center for Astrophysics \textbar \ Harvard \& Smithsonian, 60 Garden Street, Cambridge, MA 02138, USA}

\author[0000-0002-5741-3047]{David R. Ciardi}  
\affiliation{Caltech/IPAC, 1200 E. California Blvd. Pasadena, CA 91125, USA}

\author[0000-0001-6637-5401]{Allyson Bieryla}  
\affiliation{Center for Astrophysics \textbar \ Harvard \& Smithsonian, 60 Garden Street, Cambridge, MA 02138, USA}

\author{Alexandr A. Belinski} 
\affiliation{Sternberg Astronomical Institute, Lomonosov Moscow State University, 119992, Moscow, Russia, Universitetskii prospekt, 13}


\author[0000-0002-9718-3266]{Christopher A. Watson}  
\affiliation{Astrophysics Research Centre, School of Mathematics and Physics,
Queen's University Belfast, Belfast, BT7 1NN, UK}

\author[0000-0001-6763-6562]{Roland Vanderspek}  
\affiliation{Department of Earth, Atmospheric and Planetary Sciences, Massachusetts Institute of Technology, Cambridge, MA 02139, USA}
\affiliation{Kavli Institute for Astrophysics and Space Research, Massachusetts Institute of Technology, Cambridge, MA 02139, USA}

\author{St\'ephane Udry}  
\affiliation{Observatoire Astronomique de l'Universit\'e de Gen\`eve, Chemin Pegasi 51, 1290 Versoix, Switzerland}

\author[0000-0002-7504-365X]{Alessandro Sozzetti}  
\affiliation{INAF - Osservatorio Astrofisico di Torino, Strada Osservatorio 20,
  Pino Torinese (To) 10025, Italy}

\author{Damien S\'egransan}  
\affiliation{Observatoire Astronomique de l'Universit\'e de Gen\`eve, 51 chemin
  des Maillettes, 1290 Versoix, Switzerland}

\author[0000-0001-7014-1771]{Dimitar Sasselov}  
\affiliation{Center for Astrophysics \textbar \ Harvard \& Smithsonian, 60 Garden Street, Cambridge, MA 02138, USA}

\author[0000-0003-2058-6662]{George R. Ricker}  
\affiliation{Department of Physics, Massachusetts Institute of Technology, Cambridge, MA 02139, USA}
\affiliation{Kavli Institute for Astrophysics and Space Research, Massachusetts Institute of Technology, Cambridge, MA 02139, USA}

\author[0000-0002-6379-9185]{Ken Rice}  
\affiliation{SUPA, Institute for Astronomy, University of Edinburgh, Blackford
  Hill, Edinburgh, EH9 3HJ, Scotland, UK}
\affiliation{Centre for Exoplanet Science, University of Edinburgh, Edinburgh, UK}

\author[0000-0003-1200-0473]{Ennio Poretti}  
\affiliation{Fundaci\'on Galileo Galilei-INAF, Rambla Jos\'e Ana Fernandez
P\'erez 7, 38712 Bre\~{n}a Baja, TF, Spain}
\affiliation{INAF-Osservatorio Astronomico di Brera, via E. Bianchi 46, 23807
Merate (LC), Italy}

\author{Giampaolo Piotto}  
\affiliation{Dip. di Fisica e Astronomia Galileo Galilei - Universit\`a di
Padova, Vicolo dell'Osservatorio 2, 35122, Padova, Italy}

\author{Francesco Pepe}  
\affiliation{Observatoire Astronomique de l'Universit\'e de Gen\`eve, 51 chemin
  des Maillettes, 1290 Versoix, Switzerland}

\author[0000-0002-1742-7735]{Emilio Molinari}  
\affiliation{INAF - Osservatorio Astronomico di Cagliari, via della Scienza 5,
09047, Selargius, Italy}

\author[0000-0002-9900-4751]{Giuseppina Micela}  
\affiliation{INAF - Osservatorio Astronomico di Palermo, Piazza del Parlamento
1, I-90134 Palermo, Italy}

\author{Michel Mayor}  
\affiliation{Observatoire Astronomique de l'Universit\'e de Gen\`eve, 51 chemin
  des Maillettes, 1290 Versoix, Switzerland}

\author{Christophe Lovis}  
\affiliation{Observatoire Astronomique de l'Universit\'e de Gen\`eve, 51 chemin
  des Maillettes, 1290 Versoix, Switzerland}

\author[0000-0003-3204-8183]{Mercedes L\'opez-Morales}  
\affiliation{Center for Astrophysics \textbar \ Harvard \& Smithsonian, 60 Garden Street, Cambridge, MA 02138, USA}

\author[0000-0002-4715-9460]{Jon M. Jenkins}  
\affiliation{NASA Ames Research Center, Moffett Field, CA, 94035, USA}

\author[0000-0002-2482-0180]{Zahra Essack} 
\affiliation{Department of Earth, Atmospheric and Planetary Sciences, Massachusetts Institute of Technology, Cambridge, MA 02139, USA}
\affiliation{Kavli Institute for Astrophysics and Space Research, Massachusetts Institute of Technology, Cambridge, MA 02139, USA}

\author{Xavier Dumusque}  
\affiliation{Observatoire Astronomique de l'Universit\'e de Gen\`eve, 51 chemin
  des Maillettes, 1290 Versoix, Switzerland}

\author{John P. Doty} 
\affiliation{Noqsi Aerospace Ltd., 15 Blanchard Avenue, Billerica, MA 01821, USA}

\author[0000-0001-8020-7121]{Knicole D. Col\'{o}n} 
\affiliation{NASA Goddard Space Flight Center, Greenbelt, MD 20771, USA}     

\author[0000-0002-8863-7828]{Andrew Collier Cameron}  
\affiliation{School of Physics and Astronomy, University of St Andrews, North
Haugh, St Andrews, Fife, KY16 9SS, UK}

\author{Lars A. Buchhave}  
\affiliation{DTU Space, National Space Institute, Technical University of 
Denmark, Elektrovej 328, DK-2800 Kgs. Lyngby, Denmark}

%% file: stellar_table.tex
\begin{deluxetable}{lcc}
\label{tab:stellar_params}
\tabletypesize{\small}
\tablecaption{\name{} stellar parameters \label{tab:star}}
\tablewidth{0pt}
\tablehead{\colhead{Parameter} & \colhead{Value} & \colhead{Refs}}
\startdata 
\multicolumn{3}{c}{\emph{TOI-1695, TIC 422756130, 2MASS J01274094+7217472,}} \\ 
\multicolumn{3}{c}{\emph{Gaia DR2 534988616816537728}} \\ 
\multicolumn{3}{c}{\underline{Astrometry}} \\
Right ascension (J2016.0), $\alpha$ & 1:27:41.22 & 1 \\ 
Declination (J2016.0), $\delta$ & +72:17:47.83 & 1 \\ 
RA proper motion,  & $71.63\pm 0.01$ & 1 \\
$\mu_{\alpha}$ [mas yr$^{-1}$] && \\
Dec proper motion, & $40.45\pm 0.02$ & 1 \\
$\mu_{\delta}$ [mas yr$^{-1}$] && \\
Parallax, $\pi$ [mas] & $22.226\pm 0.014$ & 1 \\
Distance, $d$ [pc] & $44.993\pm 0.028$ & 2, 3 \\
\multicolumn{3}{c}{\underline{(Uncontaminated) Photometry}} \\
NUV$_{\text{GALEX}}$ & $23.99 \pm 2.18$ & 4 \\ 
$G_{\text{BP}}$ & $13.3280\pm 0.0300$ & 1 \\
$G$ & $12.1364\pm 0.0300$ & 1 \\
$G_{\text{RP}}$ & $11.0688\pm 0.0300$ & 1 \\
$T$ & $11.0294 \pm 0.0074$ & 5 \\
$J$ & $9.640\pm 0.024$ & 6 \\ 
$H$ & $8.984\pm 0.028$ & 6 \\ 
$K_s$ & $8.818\pm 0.021$ & 6 \\ 
$W1$ & $8.684\pm 0.024$ & 7 \\ 
$W2$ & $8.61\pm 0.02$ & 7 \\ 
$W3$ & $8.511\pm 0.027$ & 7 \\ 
$W4$ & $8.395001\pm 0.293$ & 7 \\ 
\multicolumn{3}{c}{\underline{Stellar Parameters}} \\
Spectral type & M1V & 8 \\
$M_{K_s}$ & $8.818\pm 0.021$ & 3 \\ 
Surface gravity, $\log{g}$ [dex] & $4.725^{+0.027}_{-0.026}$ & 3 \\ 
Metallicity, [Fe/H] [dex] & $0.0 \pm 0.5$ & 3 \\  
Effective temperature, & $3690\pm 50$ & 3 \\
\teff{} [K] && \\
Stellar radius, $R_s$ [$R_{\odot}$] & $0.515\pm 0.015$ & 3 \\ 
Stellar mass, $M_s$ [$M_{\odot}$] & $0.513\pm {0.012}$ & 3 \\
Stellar density, $\rho_s$ [g cm$^{-3}$] & $5.30^{+0.50}_{-0.45}$ & 3 \\
Stellar luminosity, $L_s$ [$L_{\odot}$] & $0.0443^{+0.0037}_{-0.0035}$ & 3 \\
\vspace{-0.15cm} Projected rotation velocity, && \\ \vspace{-0.25cm}
& $<1.3$ & 3 \\
\vsini{} [km s$^{-1}$] && \\
\logrhk{} & $-4.74\pm 0.41$ & 3 \\
Rotation period, \Prot{} [days] & $47.7\pm 2.2$ & 3
\enddata
\tablecomments{\textbf{References:}
  1) \cite{Gaia_DR3} 
  2) \cite{Bailer-Jones_2018}
  3) this work
  4) \cite{Bianchi_2017}
  5) \cite{Stassun_2019}
  6) \cite{Cutri_2003}
  7) \cite{Cutri_2021}
  8) \cite{Pecaut_2013}
}
\end{deluxetable}

%% file: point_estimates_table.tex
\begin{deluxetable}{lc}
\tabletypesize{\small}
\tablecaption{Point Estimates of the TOI-1695 Model Parameters \label{tab:point_estimates}}
\tablewidth{0pt}
\tablehead{\colhead{Parameter} & \colhead{Value}}
\startdata 
\multicolumn{2}{c}{\underline{Transit Parameters}} \\
Baseline flux, $f_{0, TESS}$ [ppt] & 0.017$^{+0.042}_{-0.040}$ \\
Limb-darkening coefficient, $q_1$ & 0.32$^{+0.53}_{-0.48}$ \\ 
Limb-darkening coefficient, $q_2$ & 0.33$^{+0.51}_{-0.48}$ \\ 
ln$\rho$ [days] & 1.00$^{+0.12}_{-0.11}$ \\
ln$\sigma_{TESS}$ & -0.91$^{+0.08}_{-0.07}$ \\
\multicolumn{2}{c}{\underline{RV Parameters}} \\
\Prot{} [days] & 48.9$^{+2.2}_{-2.7}$ \\
$\sigma_{rot}$ & 5.83$^{+0.94}_{-0.92}$ \\
ln$Q_0$ & 1.4$^{+1.0}_{-0.9}$ \\
ln$dQ$ & 0.049$^{+2.026}_{-1.984}$ \\
$f$ & 0.64$^{+0.25}_{-0.31}$ \\
RV offset, $v_0$ [m s$^{-1}$] & -0.91$^{+0.91}_{-0.88}$ \\
Jitter, $s$ [m s$^{-1}$] & 2.32$^{+0.94}_{-0.98}$ \\
\multicolumn{2}{c}{\underline{TOI-1695 b Parameters}} \\
Orbital period, $P$ [days] & 3.1342791$^{+0.0000071}_{-0.0000063}$ \\
Time of midtransit, $T_0$ [BJD - 2,457,000] & 1791.52056$^{+0.00098}_{-0.00111}$  \\
Transit duration, $D$ [hr] & 1.311$^{+0.324}_{-0.189}$ \\
Transit depth, $\delta$ [ppt] & 1.236 $^{+0.083}_{-0.081}$ \\
Semi-major axis, $a$ [au] & 0.033548$^{+0.000260}_{-0.000268}$ \\
Planet-to-star radius ratio, $R_p/R_s$ & $0.034\pm 0.002$  \\
Impact parameter, $b$ & 0.69$^{+0.11}_{-0.31}$ \\
Inclination, $i$ [deg] & 87.2$^{+1.3}_{-0.5}$ \\
Eccentricity, $e$ & $<0.097$\tablenotemark{a} \\ 
Planet radius, $R_p$ [$R_\oplus$] & 1.90$^{+0.16}_{-0.14}$ \\
RV semiamplitude, $K$ [m s$^{-1}$] & $4.39\pm 0.69$ \\
Planet mass, $M_p$ [$M_\oplus$] & $6.36\pm 1.00$ \\
Bulk density, $\rho_p$ [g cm$^{-3}$] & 5.0$^{+1.8}_{-1.3}$ \\
Surface gravity, $g_p$ [m s$^{-2}$] & 17.1$^{+4.3}_{-3.4}$ \\
Escape velocity, $v_{esc}$ [km s$^{-1}$] & $20.5\pm 1.8$ \\
Instellation, $F$ [$F_\oplus$] & $39\pm 3$ \\
Equilibrium temperature, $T_{eq}$ [K] & $698\pm 14$\tablenotemark{b} \\
\enddata
\tablenotetext{a}{95\% confidence interval}
\tablenotetext{b}{Zero albedo assumed}
\end{deluxetable}

%% file: priors_table.tex
\begin{deluxetable}{lc}
\tabletypesize{\small}
\tablecaption{TESS Light Curve and RV Model Parameter Priors
\label{tab:priors}}
\tablewidth{0pt}
\tablehead{\colhead{Parameter} & \colhead{Fiducial Model Priors}}
\startdata 
\multicolumn{2}{c}{\underline{Stellar Parameters}} \\
$M_s$ $[M_\odot]$ & $\mathcal{N}$(0.513, 0.012) \\
$R_s$ $[R_\odot]$ & $\mathcal{N}$(0.515, 0.015) \\
\multicolumn{2}{c}{\underline{Light Curve Parameters}} \\
$q_1$ & $\mathcal{U}(0, 1)$\tablenotemark{a} \\
$q_2$ & $\mathcal{U}(0, 1)$\tablenotemark{a} \\
ln$\rho$ [days] & $\mathcal{U}(2, 20)$ \\ 
ln$\sigma_{TESS}$ & $\mathcal{N}$(ln(std($\mathbf{f}_{TESS}$)), 10)\tablenotemark{b} \\ 
ln$s_{TESS}$ & $\mathcal{N}$(ln(std($\mathbf{f}_{TESS}$)), 10)\tablenotemark{b} \\
$f_{0,TESS}$ & $\mathcal{N}$(0, 2) \\
\multicolumn{2}{c}{\underline{GP and RV Parameters}} \\
$P_{rot}$ [days] & $\mathcal{N}$(48, 3) \\ 
$\sigma_{rot}$ & $\mathcal{N}$(std(data), 1) \\ 
ln$Q_0$ & $|\mathcal{N}(0, 2)|$ \\ 
ln$dQ$ & $\mathcal{N}(0, 2)$ \\ 
$f$ & $\mathcal{U}(0, 1)$ \\ 
ln$s_{RV}$ [m s$^{-1}$] & $\mathcal{U}(-3, 1)$ \\
$\gamma$ [m s$^{-1}$] & $\mathcal{N}$(0, 2) \\
\multicolumn{2}{c}{\underline{Planetary Parameters}} \\
$P$ [days] & $\mathcal{N}(3.1343, 0.0001)$\tablenotemark{c} \\
$T_0$ [BJD - 2,457,000] & $\mathcal{N}(1791.518, 0.001)$\tablenotemark{c} \\
ln$K$ [m s$^{-1}$] & $\mathcal{U}(-1, 4)$ \\
ln$\delta$ [ppt] & $\mathcal{U}(-2, 2)$\tablenotemark{d} \\
$b$ & $\mathcal{U}(0, 1)$ \\
$e$ & $\mathcal{B}(0.867, 3.03)$\tablenotemark{e} \\
& half $\mathcal{N}(0, 0.32)$\tablenotemark{f}\\
$\omega$ & $\mathcal{U}(-\pi, \pi)$ \\
\enddata
\tablenotetext{a}{Reparameterization of limb darkening coefficients, $u_1$ and $u_2$ described by \cite{Kipping_2013}.}
\tablenotetext{b}{$\mathbf{f}_{TESS}$ is normalized {\tt\string PDCSAP} flux.}
\tablenotetext{c}{Priors for RV analysis were based on the TESS transit analysis. We did not impose any prior during the TESS analysis.}
\tablenotetext{d}{Informed by BLS search and fed into $R_p/R_s$.}
\tablenotetext{e}{TESS transit analysis, \cite{Kipping_2013b}}
\tablenotetext{f}{RV analysis, \cite{Eylen_2019}. Other prior values and parameterizations were explored (including $e$cos$\omega$ and $e$sin$\omega$) and revealed negligible differences in point estimates.}
\end{deluxetable}

%% file: keystone_planet_table.tex
\begin{deluxetable*}{lcccccc}
\label{tab:keystone}
\tabletypesize{\small}
\tablecaption{M dwarf keystone planet parameters}
\tablewidth{0pt}
\tablehead{Planet & Stellar Mass & Orbital Period & Planet Radius & Planet Mass & Ref. & Probability of consistency with \\
Name & [$M_{\odot}$] & [days] & [$R_{\oplus}$] & [$M_{\oplus}$] & &  an Earth-like composition, $P_i$}
\startdata 
TOI-1235 b & $0.640\pm 0.016$ & 3.445 & 1.738$^{+0.087}_{-0.076}$ & $6.91^{+0.75}_{-0.85}$ & 1 & 0.76 \\ 
TOI-776 b & $0.544\pm 0.028$ & 8.247 & $1.85\pm 0.13$ & $4.0\pm 0.9$ & 2 & 0.03 \\ 
TOI-1695 b & $0.513\pm 0.012$ & 3.134 & 1.90$^{+0.16}_{-0.14}$ & $6.36\pm 1.00$ & 3 & 0.20 \\ 
TOI-1634 b & $0.502\pm 0.014$ & 0.989 & 1.790$^{+0.080}_{-0.081}$ & $4.91^{+0.68}_{-0.70}$ & 4 & 0.09 \\ 
TOI-1685 b & $0.495\pm 0.019$ & 0.669 & 1.70$\pm 0.07$ & $3.78\pm 0.63$ & 5 & 0.05 \\ 
G 9-40 b & $0.295\pm 0.014$ & 5.746 & $1.90\pm 0.07$ & $4.00\pm 0.63$ & 6 & $7.0\times 10^{-5}$ \\
TOI-1452 b & $0.249\pm 0.008$ & 11.062 & 1.67$\pm 0.07$ & $4.82\pm 1.30$ & 7 & 0.42 \\ 
\enddata
\tablecomments{\textbf{References:}
  1) \cite{Cloutier_2020b}
  2) \cite{Luque_2021}
  3) this work
  4) \cite{Cloutier_2021a}
  5) \cite{Bluhm_2021}
  6) \cite{Luque_2022b}
  7) \cite{Cadieux_2022}
}
\end{deluxetable*}